\documentclass{aa}

\usepackage{amssymb}
\usepackage{txfonts}
\usepackage{natbib}
\usepackage{graphicx}
\usepackage{float}
\usepackage{amsmath}
\usepackage{lscape}
\usepackage{hyperref}

\begin{document}

\title{Time-integrated polarizations in GRB prompt phase via the Multi-window interpretation}

        \author{X. Wang
                \inst{1}
                \and
                M. X. Lan\inst{1}
                \and
                Q. W. Tang\inst{2}
                \and
                X. F. Wu\inst{3}
                \and
                Z. G. Dai\inst{4}
        }

        \institute{Center for Theoretical Physics and College of Physics, Jilin University,
                 Changchun 130012, China\\
                \email{lanmixiang@jlu.edu.cn}
                \and
                 Department of Physics, School of Physics and Materials Science, Nanchang University,
                 Nanchang, 330031, China\\
                \email{qwtang@ncu.edu.cn}
                \and
                 Purple Mountain Observatory, Chinese Academy of Sciences,
                 Nanjing, 210023, China\\
                \and
                 Department of Astronomy, School of Physical Sciences, University of Science and Technology of China,
                 Hefei, 230026, China\\
        }

\abstract
{}
{The multi-window observations, including the light curve and the evolutions of the spectral peak energy ($E_p$), the polarization degree (PD) and the polarization angle (PA), are used to infer the model parameters to predict the time-integrated PD in gamma-ray burst (GRB) prompt phase.}
{We select 23 GRBs co-detected by Fermi/GBM and polarization detectors (i.e., GAP, POLAR and AstroSat). In our multi-window fitting, the light curve, $E_p$ curve, PD curve and PA curve are interpreted simultaneously under the synchrotron radiation model in ordered magnetic fields  (i.e., the aligned-fields case and the toroidal-fields case).}
{For the bursts with abrupt PA rotations, the predicted time-integrated PD of the aligned-fields case roughly matches the corresponding observed best fit value, while it is higher for the toroidal-fields case. For the bursts without abrupt PA rotation(s), the predicted PDs of the aligned-fields case and the toroidal-fields case are comparable and could interpret the observational data equally well. For GRB 170206A, its observed time-resolved and time-integrated PDs are comparable and both smaller than our predicted upper limits in ordered magnetic fields. So mixed magnetic fields, i.e., the magnetic fields with both ordered and random components, should be reside in the radiation regions of this burst. Except 1 out of the total 23 bursts, the predicted time-integrated PDs, which are around $\sim44\%$ for the aligned-fields case and around $49\%$ for the toroidal-fields case, are consistent with the corresponding observed values. Therefore, consistent with the former study, the models with synchrotron radiation in ordered magnetic fields could interpret most of the current polrization data within $1\sigma$ error bar.}
{}
\maketitle

\section{Introduction}\label{intro}

Although there had been decades after gamma-ray burst (GRB), which is the most catastrophic event  in the universe, was first discovered, their origin is still indebate. The light curves of the GRBs show rich diversity, which adds the difficulty in finding their physical origin. However, the spectrum of the GRBs is often described by the Band function \citep{1993BATSE}, which is a broken power law, smoothly jointed at $(\alpha_B-\beta_B)E_p/(2+\alpha_B)$, where $E_p$ is the peak energy of the $\nu F_\nu$ spectrum and $\alpha_B$ ($\beta_B$) is the low-energy (high-energy) spectral index of the photon number flux. Observationally, there are mainly two kinds of the peak energy evolution patterns, i.e., the hard-to-soft and intensity-tracking patterns \citep{LK1996, Ford1995, Kaneko2006, Lu2010, Lu2012, Lu2018}. And the origin of the Band function is also mysterious.

Three popular models are involved to interpret the observations of GRB prompt phase \citep{1992ApJ...395L..83N, PX1994,  1994ApJ...430L..93R, 1994MNRAS.270..480T, 2000ApJ...529..146E, 2000ApJ...530..292M, 2008A&A...480..305G, 2011ApJ...726...90Z}. In the scenario of the internal shock and the magnetic reconnection models, the radiation mechanisms are both the synchrotron emission \citep{Daigne2009, Lan2021b, 2008A&A...480..305G, 2011ApJ...726...90Z}. However, it is the multiple inverse-Compton scatterings in the high-energy gamma-rays for the photosphere model \citep{1994MNRAS.270..480T, Lundman_2018, Parsotan_2020}. Since all three models could explain the observed light curves and energy spectra in GRB prompt phase, these models and hence their corresponding radiation mechanism can not be distinguished by the traditional observations. Polarization, as a new window, is very sensitive to the radiation mechanism, magnetic field configuration (MFC) in the radiation region and the geometry \citep{Waxman_2003, Granot_2003, Rossi_2004, Lazzati_2006, Toma_2009, Lan_2021, Gill_2021, Guan_2023, SL2024}. Therefore, it can be used as a powerful tool to infer these properties of the radiating sources.

Because of the depolarization due to the multiple scatterings, the predicted polarization degree (PD) is roughly 0 in the gamma-ray band for the photosphere model \citep{Lundman_2018, Parsotan_2020, Parsotan_2022}. However, the predicted time-integrated PD can be as high as $(40-50)\%$ for the synchrotron model in an ordered magnetic field \citep{Toma_2009, Gill2020, Gill_2021,  Guan_2023, SL2024}. If the polarization angle (PA) rotates with time (or energy) and/or the magnetic field is mixed \footnote{There are both the ordered component and the random component of the magnetic field \citep{Lan_2019}.} in the radiation region, there would be a reduction of the time- and energy-integrated PD. So the predicted time-integrated PD can range from $0\%$ to $50\%$ for the synchrotron models \citep{Lan_2019, SL2024}. Therefore, the precise PD observations are important in distinguishing the models and hence the corresponding radiation mechanisms during GRB prompt phase. If the gamma-ray emission in GRB prompt phase are unpolarized, the photosphere model survives and the synchrotron model with a totally random magnetic field could also interpret such observations. However, if the PD of the gamma-ray radiation are non-zero, the photosphere model is rejected. 

The Gamma-Ray Burst Polarimeter (GAP) is a polarization detector to measure the linear polarizations in GRB prompt phase in the energy range of (70 keV, 300 keV) \citep{Yonetoku_2006, Murakami_2010, Yonetoku_2011b}. POLAR is also a dedicated GRB polarimeter and the linear polarizations were measured in the energy range of (50 keV, 500 keV) \citep{Produit_2018}. The Cadmium Zinc Telluride Imager on board AstroSat is an in operation GRB polarization detector and its polarization measurement is done in the energy range of (100 keV, 600 keV) \citep{Chattopadhyay_2022}. There should be sufficient photons to do polarization analysis, most of the current polarization observations are the time- and energy-integrated results \citep{Yonetoku_2011,Yonetoku_2012,Zhang_2019,Kole_2020,Chattopadhyay_2022}. The time-resolved observations are relatively rare and the time-resolved polarization analysis had been done in only 4 GRBs  \citep{Yonetoku_2011,Zhang_2019,Burgess_2019,Kole_2020,Sharma_2019,Chattopadhyay_2022}. 

The number of the bursts detected by the polarization detector GAP is 3 and the averaged best-fit value of the time-integrated PD is $60\%$ for GAP bursts \citep{Yonetoku_2011,Yonetoku_2012}. The observed time-integrated PDs of 15 bursts among the total 20 bursts detected by AstroSat are the upper limits and only 5 bursts are reported with concrete PD values \citep{Chattopadhyay_2022}. The averaged best-fit value of the time-integrated PD of these 5 bursts is $57.83\%$. There are 14 GRBs with reported polarization detections by the POLAR \citep{Zhang_2019,Kole_2020} and the averaged best-fit value of the time-integrated PD is $17.79\%$. The smallest lower limit of the detected time-integrated PD within $1\sigma$ confidence level is $16\%$ for GAP bursts and the highest upper limit of the detected PD within $1\sigma$ is $84\%$ for POLAR bursts. Since the observational errors are rather large and confidence levels (usually within $1\sigma$) are relatively low, the current data are in general not contradict. 

Recently, time-integrated PDs in GRB prompt phase had been studied by \cite{Guan_2023}. They used the observed parameters of the energy spectra to estimate the time-integrated PD, in which the time-resolved information was missed. However, up till now, there are 4 GRBs with time-resolved polarization observations. Among these 4 GRBs, there are 3 bursts with PA rotating abruptly at least once during the burst, which would lead to a reduction of the time-integrated PD. And also the variations of the parameters on the equal arrival time surface (EATS) would also impact the final time-integrated PDs \citep{SL2024}. Therefore, the time-evolution effect should not be neglected in deducing the time-integrated PDs. The parameters are taken as their typical or fiducial values and vary in reasonable ranges to predicted the time-integrated PDs in \cite{SL2024}. Actually, in addition to the polarization observations, both the observed light curves and energy spectra would put limits on the model parameters. Therefore, in order to predict more precise time-integrated PDs in GRB prompt phase, multi-window interpretations (i.e., the simultaneous interpretation/fitting of the light curve, $E_p$ curve, PD curve and the PA curve) are needed.

In this paper, the time-integrated PDs in GRB prompt phase are studied via the time-resolved multi-window fitting. The EATS efffect is considered and the time-resolved information is included. The observed PA rotations are recovered during the bursts. More important, most of the key model parameters are inferred from the fitting of the light curves, the energy spectra and the polarization curves. The predicted time-integrated PD for each burst is customized. In Section 2, the data processing is presented. In Section 3, we describe our model. In Section 4, our numerical results are presented. Finally, we give our conclusions and discussion in Section 5.

\section{Data Reduction}

The light curves and $E_p$ curves are obtained from the Fermi/GBM data through the following procedure. $RMFit$ v4.3.2 is employed to perform data analysis~\citep{2014ascl.soft09011G}. Observational database are downloaded in the official web site of $Fermi$ Gamma-ray Burst Monitor (GBM)~\footnote{https://heasarc.gsfc.nasa.gov/W3Browse/fermi/fermigbrst.html}. For each GRB, we select two sodium iodide (NaI) detectors that are most close to the GRB center as well as one bismuth germanate (BGO) detector. GBM time-tagged events (TTE) with 2 $\mu$s precision are used in the data reduction. Photon energy is selected from 8 keV to 1000 keV for NaI detector, while from 200 keV to 40 MeV for BGO detector. We select two intervals as the background intervals and employ a polynomial model with 0-3 order to fit the background. After that, we perform the time-averaged and time-resolved spectral fitting using a chi-square statistics. For the time-averaged spectrum, the GBM $T_{90}$ interval in the 50-300 keV energy band is selected as the analyzed duration~\citep{vonKienlin_2020}. For the time-resolved spectral analyses, we derive the time bins by setting a signal to noise ratio (SNR) between 12 and 100 among GBM $T_{90}$. A typical Band function is employed to fit all spectra~\citep{1993BATSE}. Spectral parameters (e.g., $E_p$) and photon flux (between 8 keV and 40 MeV) of individual spectrum are calculated in the best-fit result.

The data of the polarizations are taken from the published papers \citep{Yonetoku_2011, Yonetoku_2012, Zhang_2019, Kole_2020, Chattopadhyay_2022}. The energy ranges for the polarization data are (70 keV, 300 keV), (50 keV, 500 keV) and (100 keV, 600 keV) for the bursts detected by the GAP, POLAR and AstroSat, respectively \citep{Yonetoku_2011, Yonetoku_2012, Zhang_2019, Kole_2020, Chattopadhyay_2022}. In our sample, there are total 23 bursts, which are co-detected by Fermi/GBM and polarization detectors. The main information of these bursts are shown in Table \ref{table:integrated}. The confidence level for the errors of the observed time-integrated PD ($PD_{obs}$) is $1\sigma$. It is $2\sigma$ for the upper limits of the $PD_{obs}$. The confidence level for the time-integrated photon spectral indices ($\alpha_B$ and $\beta_B$) are also $1\sigma$. In our sample, the average value of the observed time-integrated PD is around $60\%$ for GAP bursts, around $50\%$ for the AstroSat bursts and around $22\%$ for the POLAR bursts. The minimum lower limit of the observed time-integrated PD within $1\sigma$ is $16\%$ for GAP bursts, it is $28.32\%$ for AstroSat bursts. The maximum upper limit of the observed time-integrated PD within $1\sigma$ is $84\%$ for POLAR bursts. So because of large observational errors, the detection results of the three detectors are in general not contradictory. In addition, the errors of the current data are roughly within the confidence level of $1\sigma$. These results may be coincide with each other in higher confidence level.

\section{The model}\label{model}

The model used here for fitting the time-resolved multi-window observations is same to that in \cite{Wang_2024}. We assume that GRB at redshift $z$ originates from the synchrotron radiation of the multiple thin emitting shells. These shells, with different physical properties, are injected discontinuously because of the activities of the GRB central engine. They would expand radially at relativistic velocities. The process of the magnetic reconnection in each individual shell would deplete the magnetic energy and provides the energy for the acceleration and radiation of the shell. The relativistic electrons would be injected isotropically into each shell with an rate of $R_{inj}$ and radiate the synchrotron photons in the magnetic fields. Here, the single-energy electrons are assumed \citep{Uhm_2018}. For ``i'' model (corresponding to the observed hard-to-soft $E_p$ evolution pattern), the variation of the electron Lorentz factor with radius is assumed to be $\gamma_{ch}(r) = \gamma_{ch}^{0} (r/r_0)^{g}$, where $\gamma_{ch}^{0}$ is the value of electron Lorentz factor at ${r_0}$. While for ``m'' model (the corresponding $E_p$ evolution pattern is intensity-tracking), we take $\gamma_{ch}(r)= \gamma_{ch}^{m}(r/r_{m})^{g}$ for $r\leq r_{m}$ and $\gamma_{ch}(r)= \gamma_{ch}^{m}(r/r_{m})^{-g}$ for $ r> r_{m}$, where $\gamma_{ch}^{m}$ is value of the electron Lorentz factor at the normalization radius $r_{m}$. Unless specified, we take $g=-0.2$ for the ``i'' model and $g=1.0$ for the ``m'' model, respectively \citep{Uhm_2018}.

All the shells will begin to radiate at radius $r_{on}$ at burst source time $t_{on}$ and stop at radius $r_{off}$. The dynamics of the shells would depend on the central engine \citep{Drenkhahn_2002}. For a perpendicular rotator \footnote{The magnetic axis of the magnetar is perpendicular to its rotational axis.}, the magnetic field in its ejecta is aligned and the reconnection of such field would lead to the bulk acceleration of the jet shell. For a black hole central engine, the magnetic field in jet is toroidal and the reconnection of the toroidal field would lead to a roughly constant-velocity jet shell at large radii. The bulk Lorentz factor of each shell then reads $\Gamma(r) = \Gamma_{0} (r/r_0 )^s$, where $\Gamma_{0}$ is the bulk Lorentz factor of the shell at the normalization radius $r_0$ and $s=1/3$ (0) for the aligned-fields (the toroidal-fields) case \citep{Drenkhahn_2002}. Meanwhile, the magnetic field in the shell decays with radius $B'(r) = B'_{0} (r/r_0)^{-b}$, where $B'_0=30$ G is the strength of the magnetic field at $r_0$ and the decay index $b$ is taken as 1.0. For simplicity, the MFC in each jet shell is assumed to be large-scale ordered. Except the MFC, the other physical properties are assumed to be independent between the emitting shells of one burst.

The time-resolved and energy-averaged flux density of the emission from a burst is 
\begin{equation}
f_\nu=\frac{\int^{\nu_2}_{\nu_1}F_{\nu}d\nu}{\int^{\nu_2}_{\nu_1}d\nu}
\end{equation}
where $h{\nu_1}=8$ keV, $h{\nu_2}=40$ MeV. The $\nu$ is the frequency and $F_{\nu}$ is the time- and energy-resolved flux density of the single burst and the corresponding expression can be found in \cite{Wang_2024}.
 
The predicted time-resolved and energy-integrated PD and PA (preliminary) of a burst with aligned fields in all the radiation shells (aligned-fields case) read,
\begin{equation}\label{eq:Pi}
PD=\frac{\sqrt{Q^2+U^2}}{F}
\end{equation}
\begin{equation}
PA_{pre}=\frac{1}{2}\arctan\left(\frac{U}{Q}\right)
\end{equation}
where $F=\int^{\nu_2}_{\nu_1}F_{\nu}d\nu$, $Q=\int^{\nu_2}_{\nu_1}Q_{\nu}d\nu$ and $U=\int^{\nu_2}_{\nu_1}U_{\nu}d\nu$ are the time-resolved and energy-integrated Stokes parameters. And $h{\nu_1}$ and $h{\nu_2}$ here are the lower- and upper-limit of the corresponding polarization detectors. And $Q_{\nu}$ and $U_{\nu}$ are the time- and energy-resolved Stokes parameters Q and U of a single burst, respectively. Their expressions can be found in \cite{Wang_2024}. If $Q>0$, the final PA ($PA$) of a burst is $PA=PA_{pre}$. If $Q<0$, the final PA is $PA=PA_{pre}+\pi/2$ when $U>0$ and is $PA=PA_{pre}-\pi/2$ when $U<0$ \citep{Lan2018}.

For a burst with toroidal fields in all its emitting shells (toroidal-fields case), because of the axial symmetry of the emission region and the reference axis selected here, the Stokes parameter $U_{\nu}$ equals to zero. The time-resolved and energy-integrated PD is defined as follows.
\begin{equation}
PD=\frac{Q}{F}
\end{equation}
When $PD$ changes its sign, the corresponding PA would change abruptly by 90°.

Finally, the time-integrated and energy-integrated PD of a burst for the aligned-fields case read as follows.
\begin{equation}\label{eq:Pi}
PD_{cal,a}=\frac{\sqrt{\bar{Q}^2+\bar{U}^2}}{\bar{F}}
\end{equation}
where $\bar{F}=\int^{T_{95}}_{T_5}Fdt/T_{90}$, $\bar{Q}=\int^{T_{95}}_{T_5}Qdt/T_{90}$, and $\bar{U}=\int^{T_{95}}_{T_5}Udt/T_{90}$ are the time-averaged and energy-integrated Stokes parameters. The accumulated flux reaches $5\%$ ($95\%$) of the total flux of the burst at the time of $T_5$ ($T_{95}$), and $T_{90}=T_{95}-T_5$. For the toroidal-fields case, the time-integrated and energy-integrated PD is
\begin{equation}
PD_{cal,t}=\frac{\bar{Q}}{\bar{F}}
\end{equation}

\section{The numerical results}\label{result}

\subsection{The time-resolved multi-window fitting results of the total 23 GRBs}\label{resolved}

The concrete fitting procedure of the time-resolved multi-window observations for a single GRB can be found in \cite{Wang_2024}. The half-opening angles of the jets for all 23 bursts are set to be the typical value of $\theta_j=0.1$ rad \citep{RE2023,Lloyd2019}. Unless specified, the shells begin to radiate at $r_{on}=10^{14}$ cm and stops at $r_{off}=3\times10^{16}$ cm. Since the GRBs with polarization detections are usually very bright, indicating on-axis observations. Because the predicted time- and energy-integrated PD is roughly independent on the observational angles for on-axis observations of the aligned-fields case \citep{SL2024, Lan_2021}, here we simply take the observational angle $\theta_V=0$ rad for the aligned-fields case. The normalization value of bulk Lorentz factor for the aligned-fields case is taken as $\Gamma_0=250$. For the toroidal-fields case, we take $\theta_V=0.06$ rad and the normalization radius $r_0=10^{15}$ cm. With the adopted parameters, the main radiation region ($1/\Gamma$ cone) is far smaller than the jet region, and the MFC within the radiation cone is roughly aligned for the toroidal-fields case. Hence the predicted time- and energy-integrated PD is roughly the theoretical upper limits for the toroidal-fields case \citep{SL2024, Lan_2021, Toma_2009}. The low- and high-energy photon spectral indices $\alpha_B$ and $\beta_B$ of each burst are set as their observed time-integrated values, which are listed in Table \ref{table:integrated}. For the bursts without redshift observation, we set $z=1$. The other parameters used in fitting can be found in Table \ref{table:parameters}. For each burst, the parameter set with the parameter $\delta$ (the orientation of the field) is for the aligned-fields case and the other is for the toroidal-fields case. Compared with the observed $E_p$ curve, the decay stage of the calculated one of GRB 170305A is very shallow with the index $g=-0.2$ for the``i'' model, so we take $g=-1.0$ for the aligned-fields case and $g=-1.5$ for the toroidal-fields case to match with the observational data.

\begin{figure*}[h]
\centering
\includegraphics[width=0.9\linewidth]{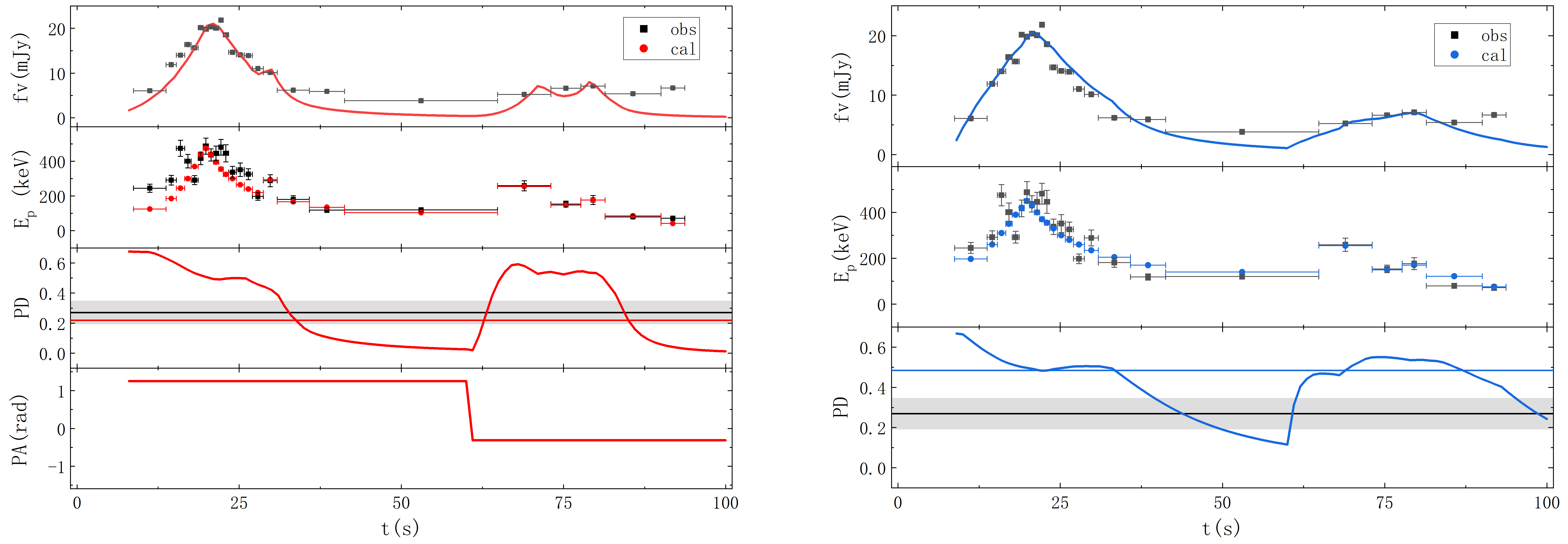}
\caption{\label{fig:100826A}Time-resolved fitting result of GRB 100826A. The four left panels show the light curve, $E_p$ curve, PD curve and PA curve in proper sequence for the aligned-fields case, while the three right panels show the light curve, the evolution of $E_p$ and the PD curve for the toroidal-fields case. The black squares show the observational data. The red circles and lines show our fitting results for the aligned-fields case, while the blue ones show the fitting result for the toroidal-fields case. The observed time-integrated PD with its $1\sigma$ error are shown as a black horizontal line with gray region, while the predicted one is shown as red (blue) horizontal line for the aligned-fields (toroidal-fields) case. Since the orientation of the GRB and the direction of the aligned field (if have) are stochastic, the predicted time-integrated PAs are not compared with the corresponding observational values for the two cases.}
\end{figure*}

\begin{figure*}[h]
\centering
\includegraphics[width=0.9\linewidth]{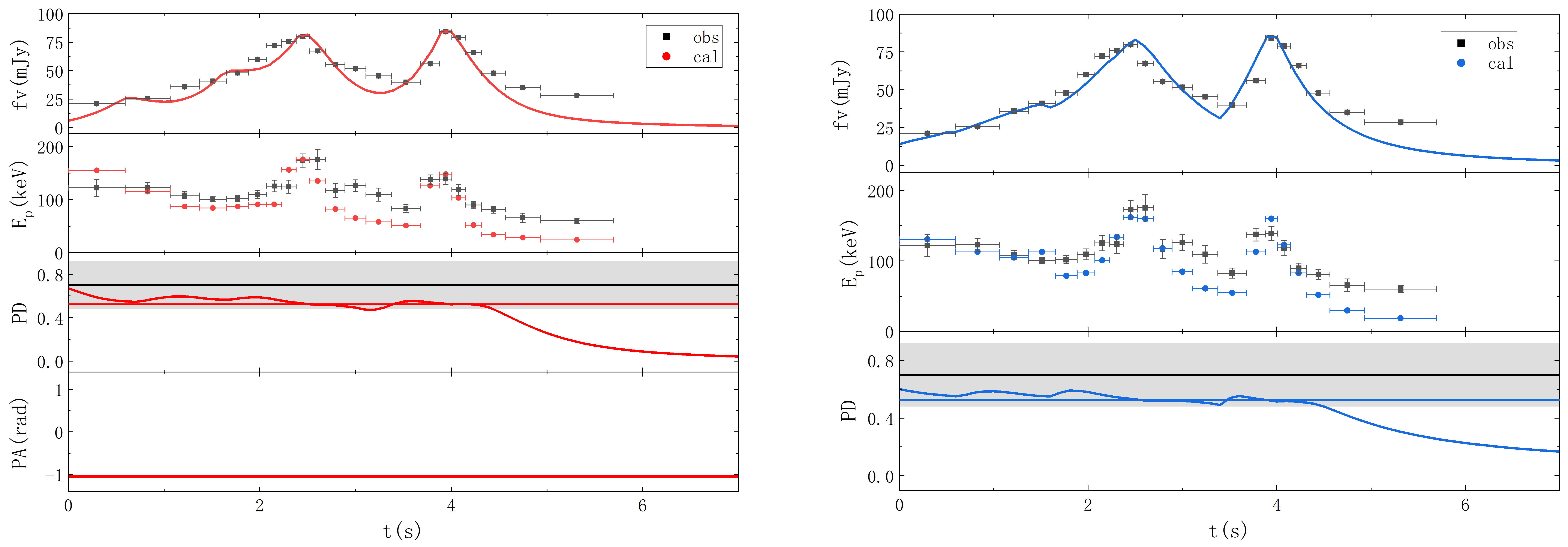}
\caption{\label{fig:110301A}Same as Figure. \ref{fig:100826A}, but for GRB 110301A.}
\end{figure*}

\begin{figure*}[h]
\centering
\includegraphics[width=0.9\linewidth]{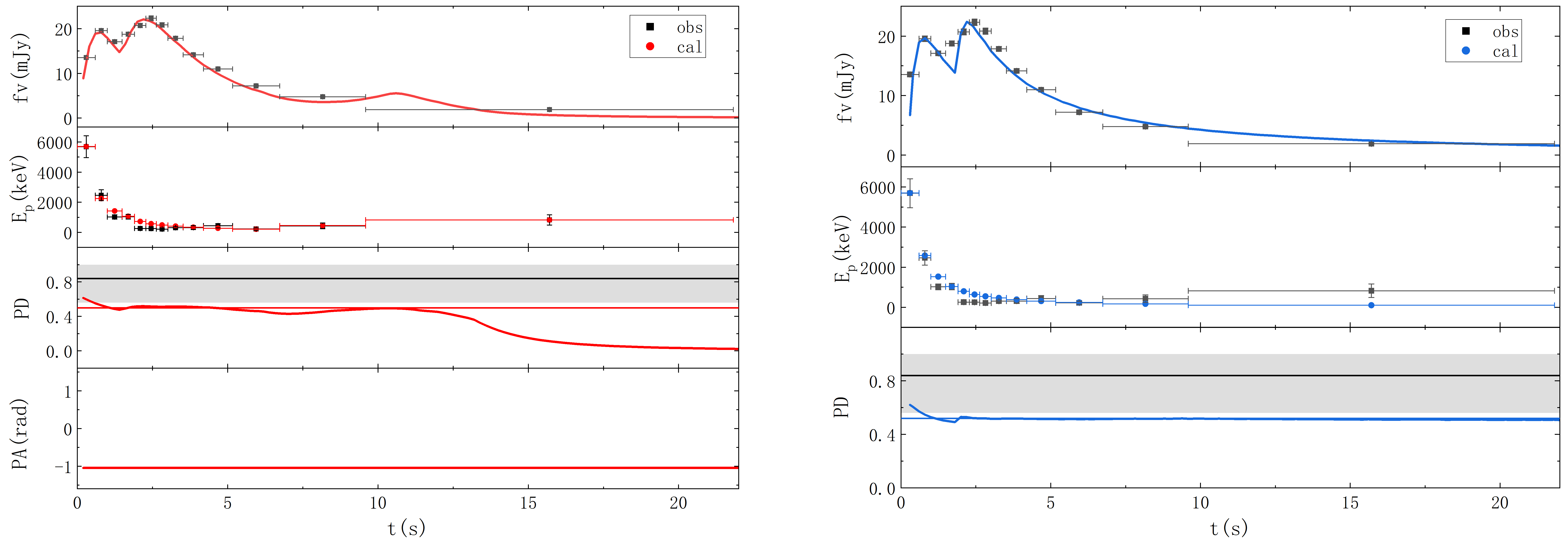}
\caption{\label{fig:110721A}Same as Figure. \ref{fig:100826A}, but for GRB 110721A.}
\end{figure*}

\begin{figure*}[h]
\centering
\includegraphics[width=0.9\linewidth]{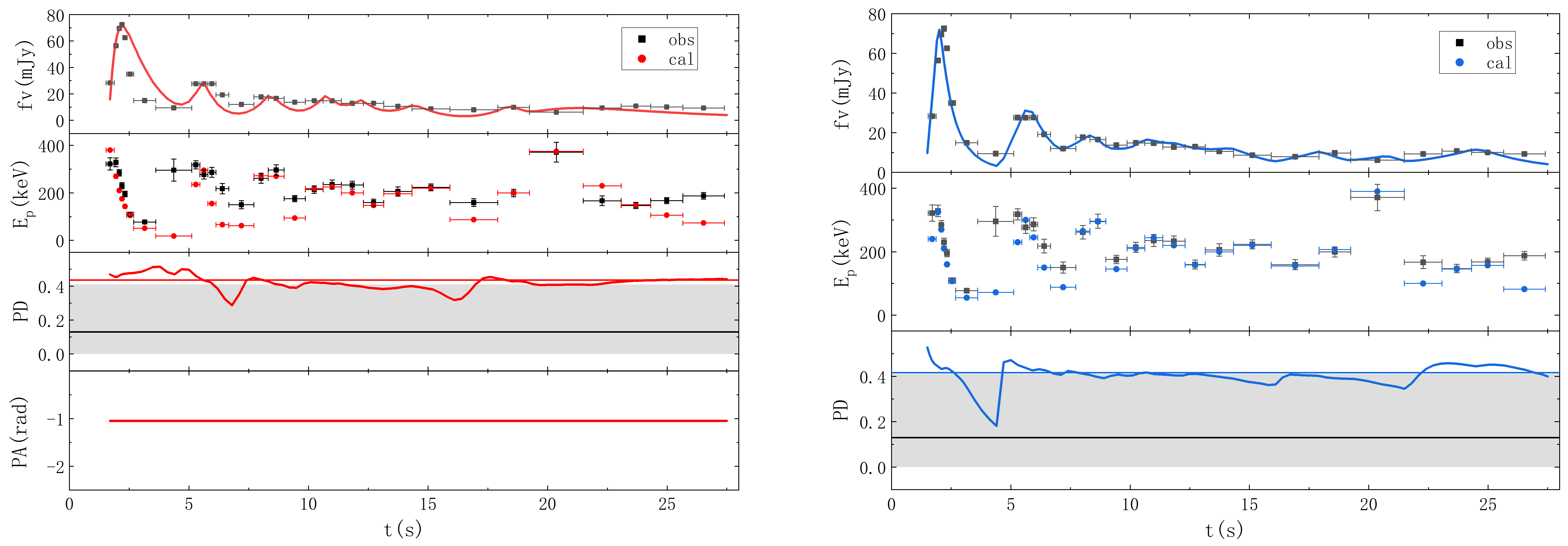}
\caption{\label{fig:161218B}Same as Figure. \ref{fig:100826A}, but for GRB 161218B.}
\end{figure*}

\begin{figure*}[h]
\centering
\includegraphics[width=0.9\linewidth]{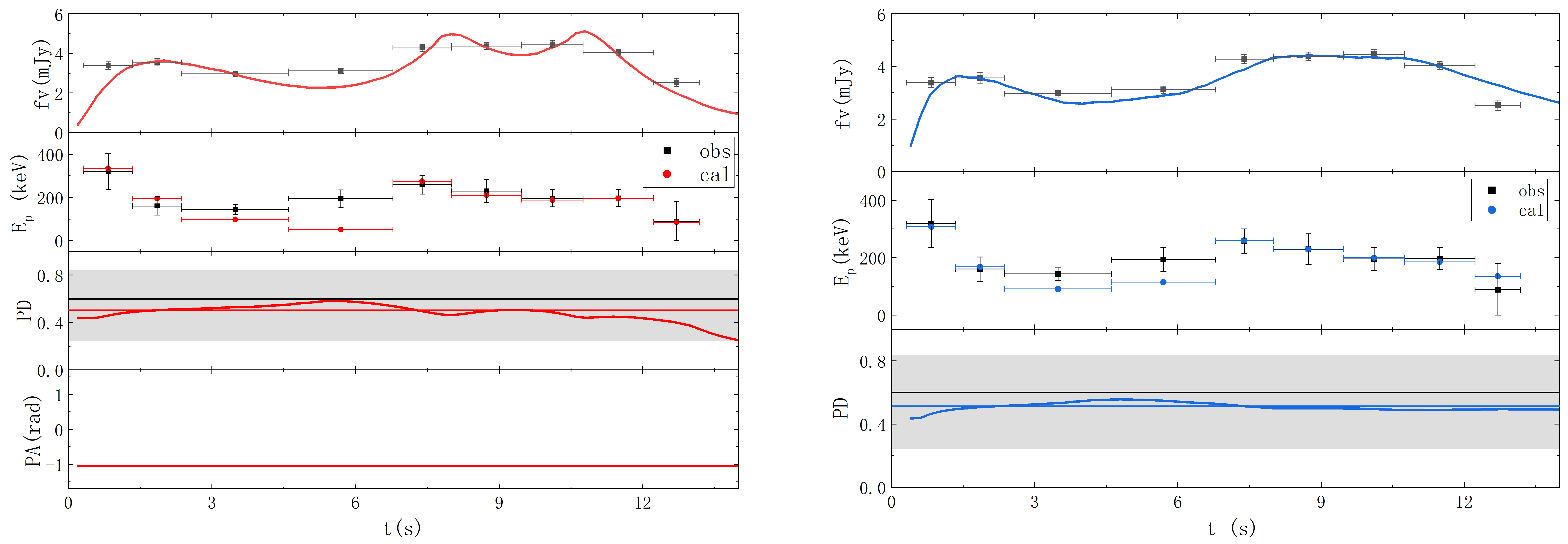}
\caption{\label{fig:170101B}Same as Figure. \ref{fig:100826A}, but for GRB 170101B.}
\end{figure*}

\begin{figure*}[h]
\centering
\includegraphics[width=0.9\linewidth]{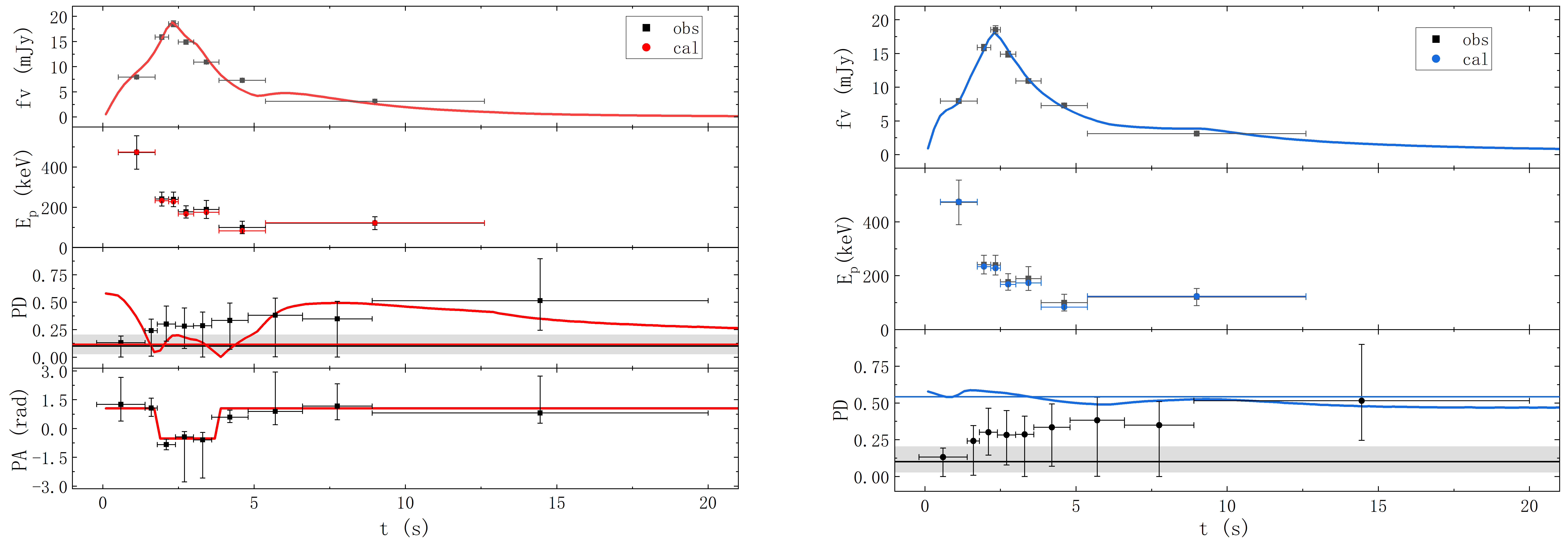}
\caption{\label{fig:170114A}Same as Figure \ref{fig:100826A}, but for GRB 170114A.}
\end{figure*}

\begin{figure*}[h]
\centering
\includegraphics[width=0.9\linewidth]{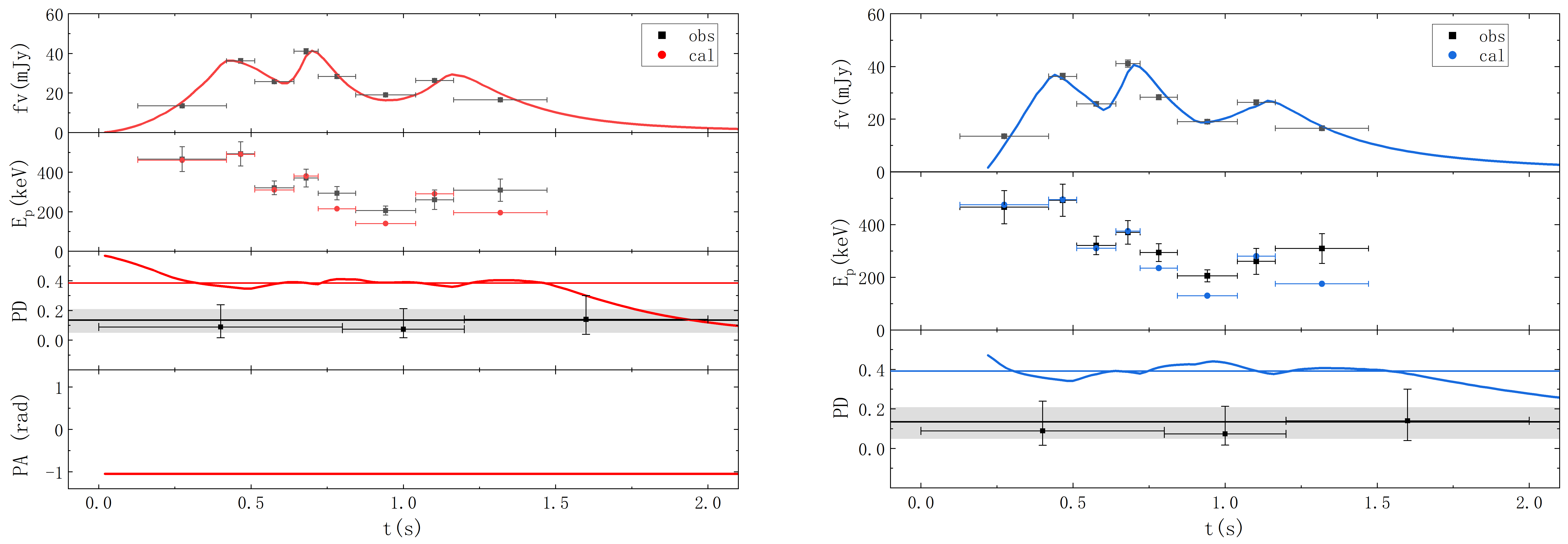}
\caption{\label{fig:170206A}Same as Figure \ref{fig:100826A}, but for GRB 170206A.}
\end{figure*}

\begin{figure*}[h]
\centering
\includegraphics[width=0.9\linewidth]{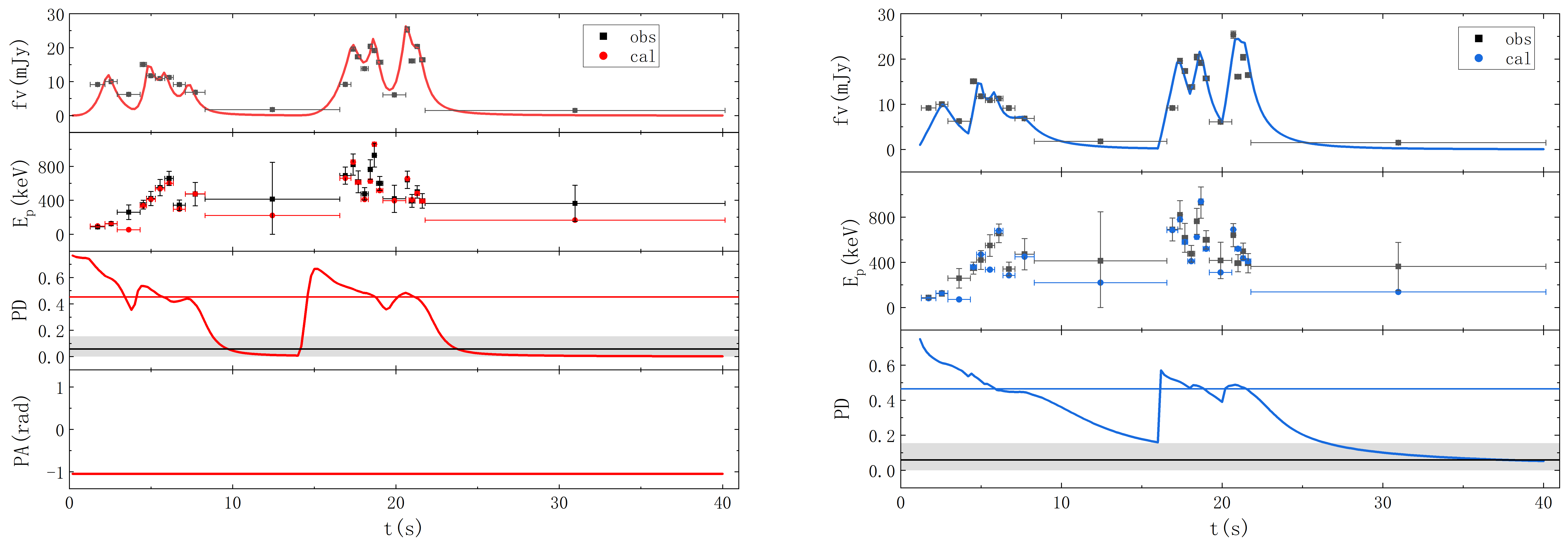}
\caption{\label{fig:170207A}Same as Figure. \ref{fig:100826A}, but for GRB 170207A.}
\end{figure*}

\begin{figure*}[h]
\centering
\includegraphics[width=0.9\linewidth]{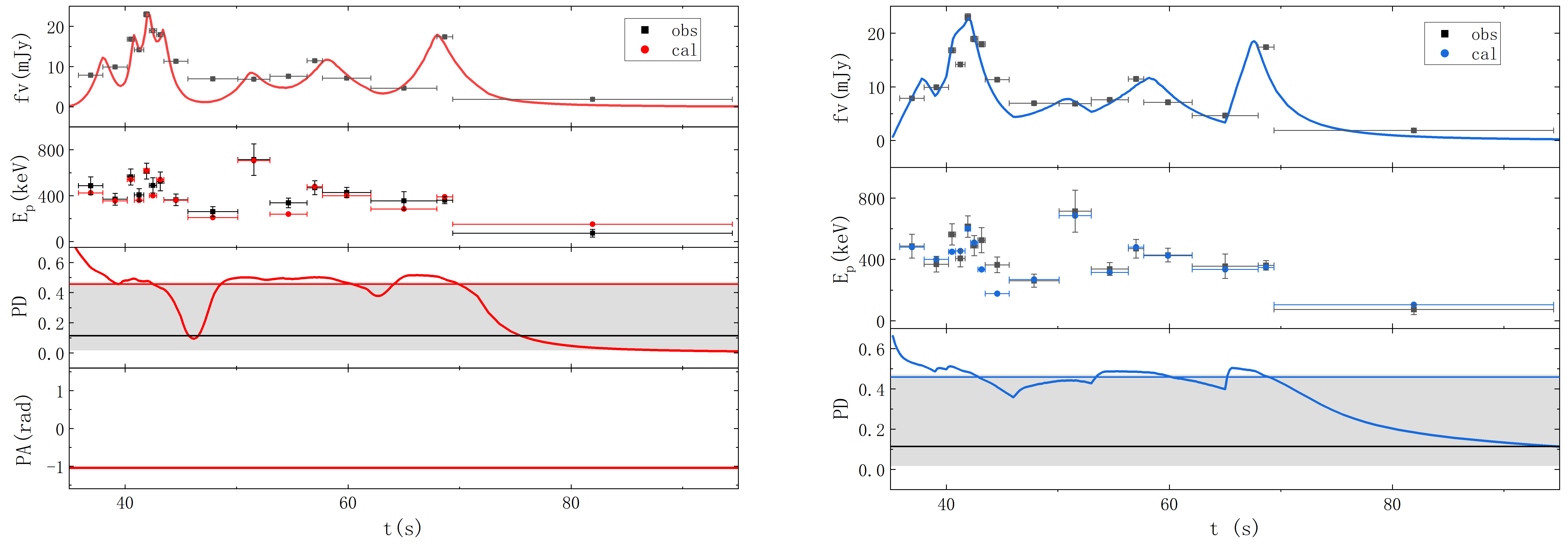}
\caption{\label{fig:170210A}Same as Figure. \ref{fig:100826A}, but for GRB 170210A.}
\end{figure*}

\begin{figure*}[h]
\centering
\includegraphics[width=0.9\linewidth]{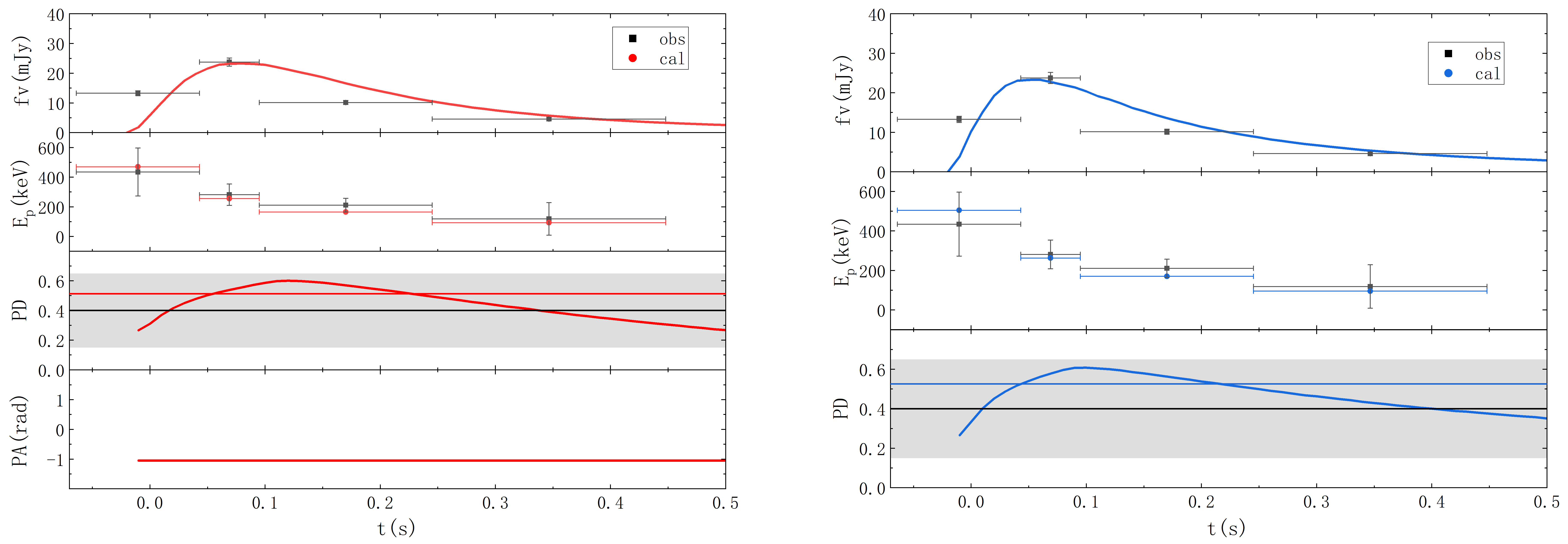}
\caption{\label{fig:170305A}Same as Figure. \ref{fig:100826A}, but for GRB 170305A.}
\end{figure*}

\begin{figure*}[h]
\centering
\includegraphics[width=0.9\linewidth]{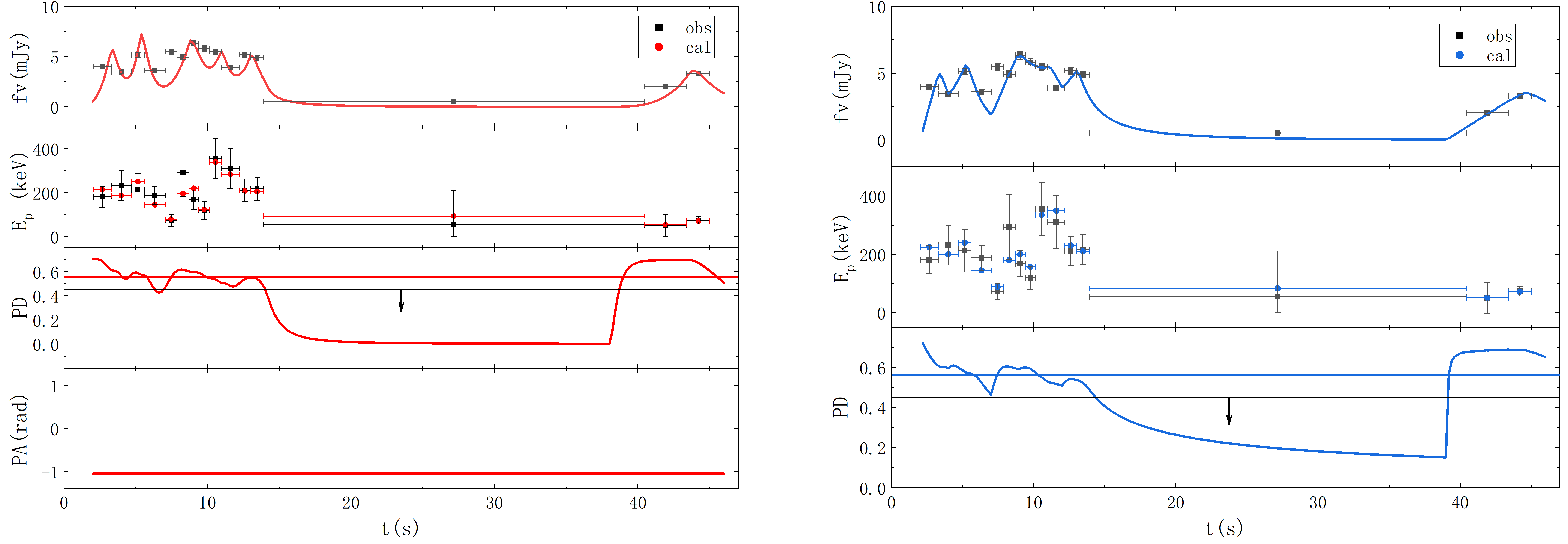}
\caption{\label{fig:160325A}Same as Figure. \ref{fig:100826A}, but for GRB 160325A.}
\end{figure*}

\begin{figure*}[h]
\centering
\includegraphics[width=0.9\linewidth]{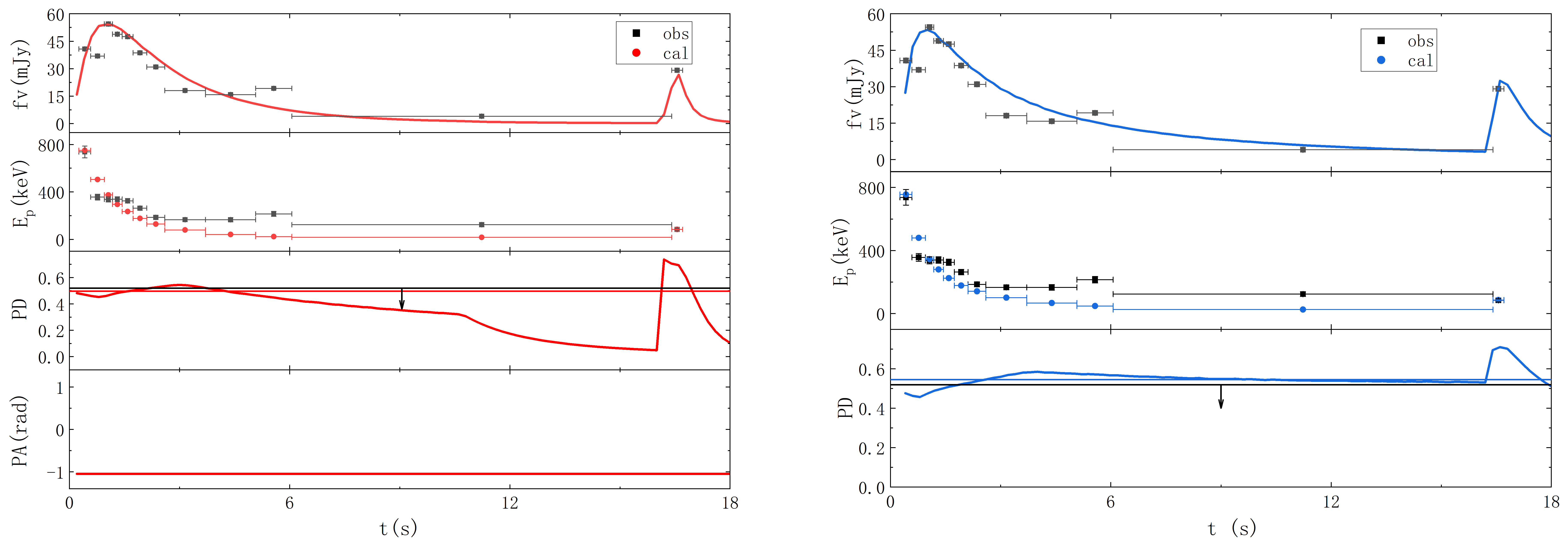}
\caption{\label{fig:160802A}Same as Figure. \ref{fig:100826A}, but for GRB 160802A.}
\end{figure*}

\begin{figure*}[h]
\centering
\includegraphics[width=0.9\linewidth]{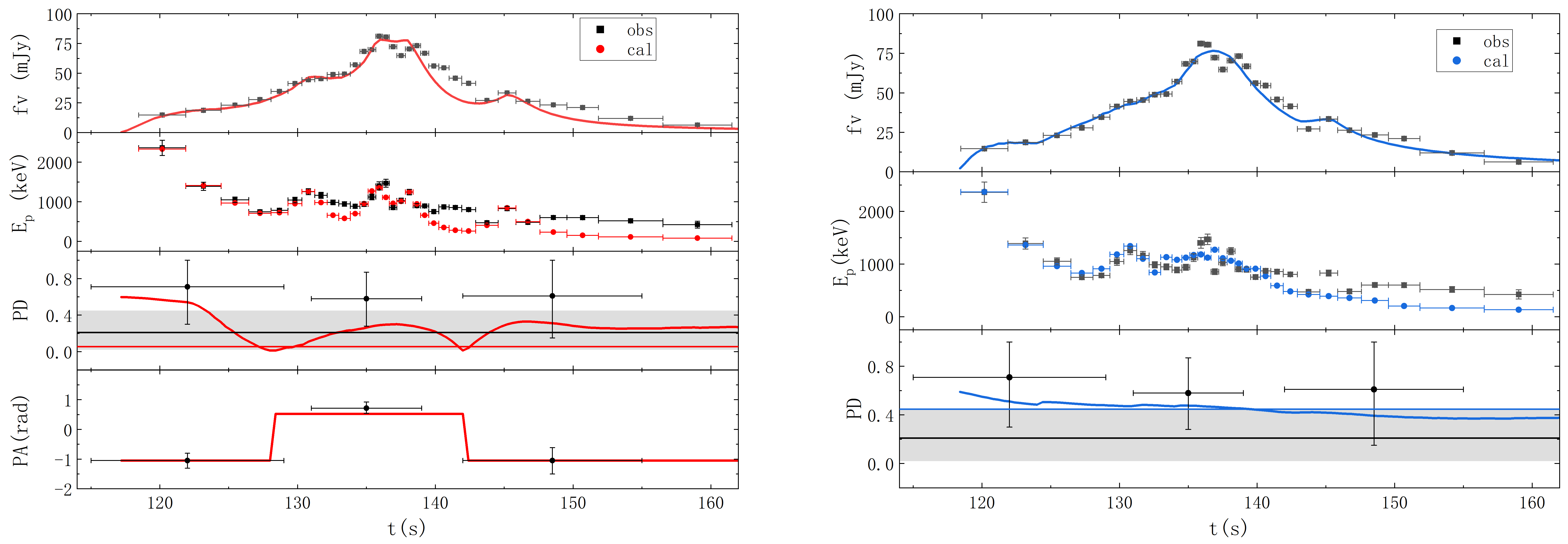}
\caption{\label{fig:160821A}Same as Figure \ref{fig:100826A}, but for GRB 160821A. The calculated energy range for the time-resolved PD and PA is (100 keV, 300 keV) as in \cite{Sharma_2019}, which is different from (100 keV, 600 keV) of the time-integrated results in Table \ref{table:integrated}.}
\end{figure*}

\begin{figure*}[h]
\centering
\includegraphics[width=0.9\linewidth]{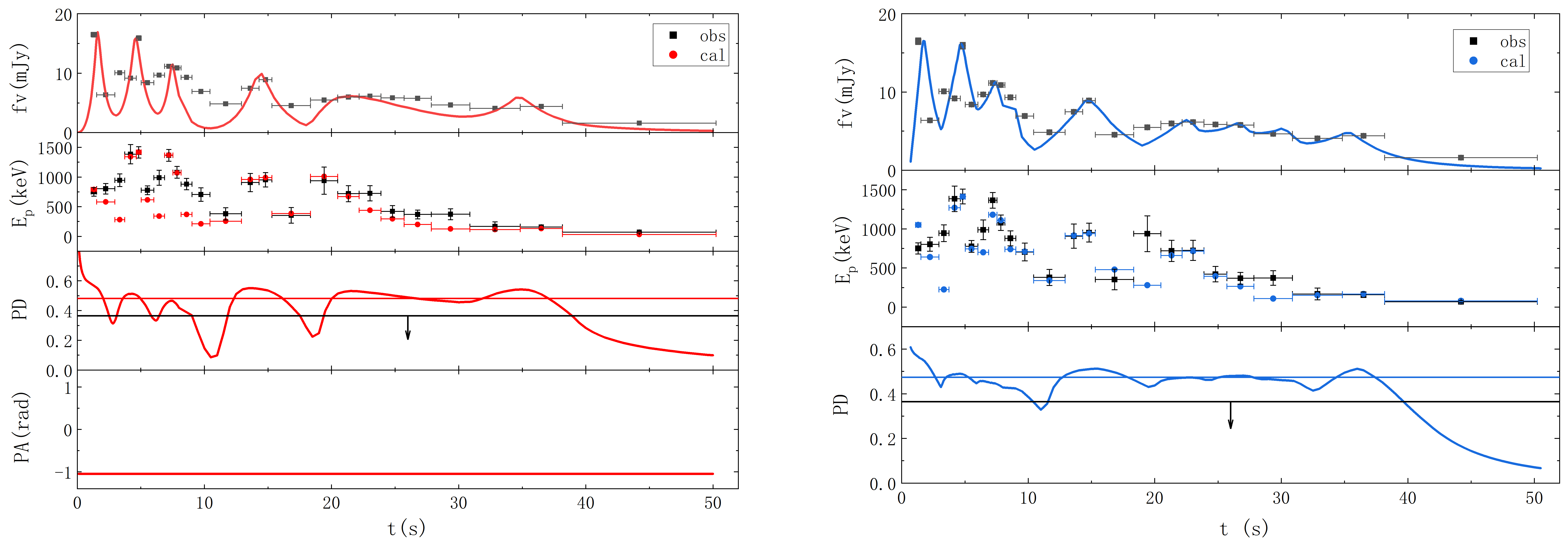}
\caption{\label{fig:170527A}Same as Figure. \ref{fig:100826A}, but for GRB 170527A.}
\end{figure*}

\begin{figure*}[h]
\centering
\includegraphics[width=0.9\linewidth]{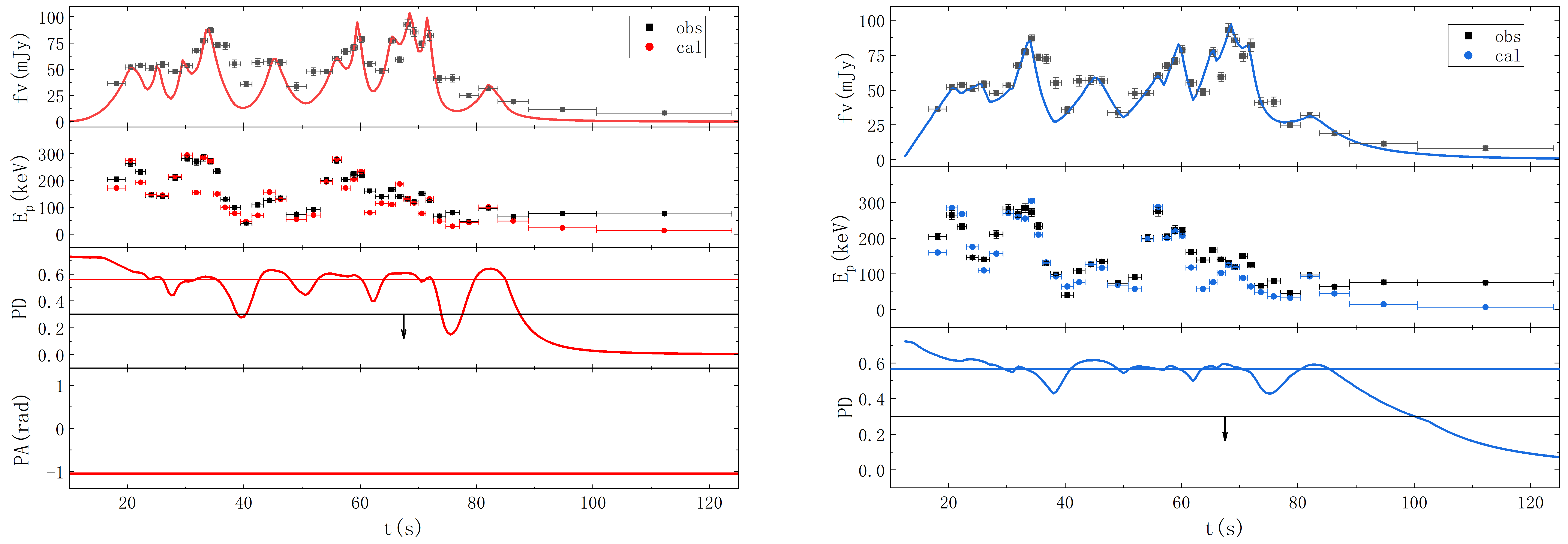}
\caption{\label{fig:171010A}Same as Figure. \ref{fig:100826A}, but for GRB 171010A.}
\end{figure*}

\begin{figure*}[h]
\centering
\includegraphics[width=0.9\linewidth]{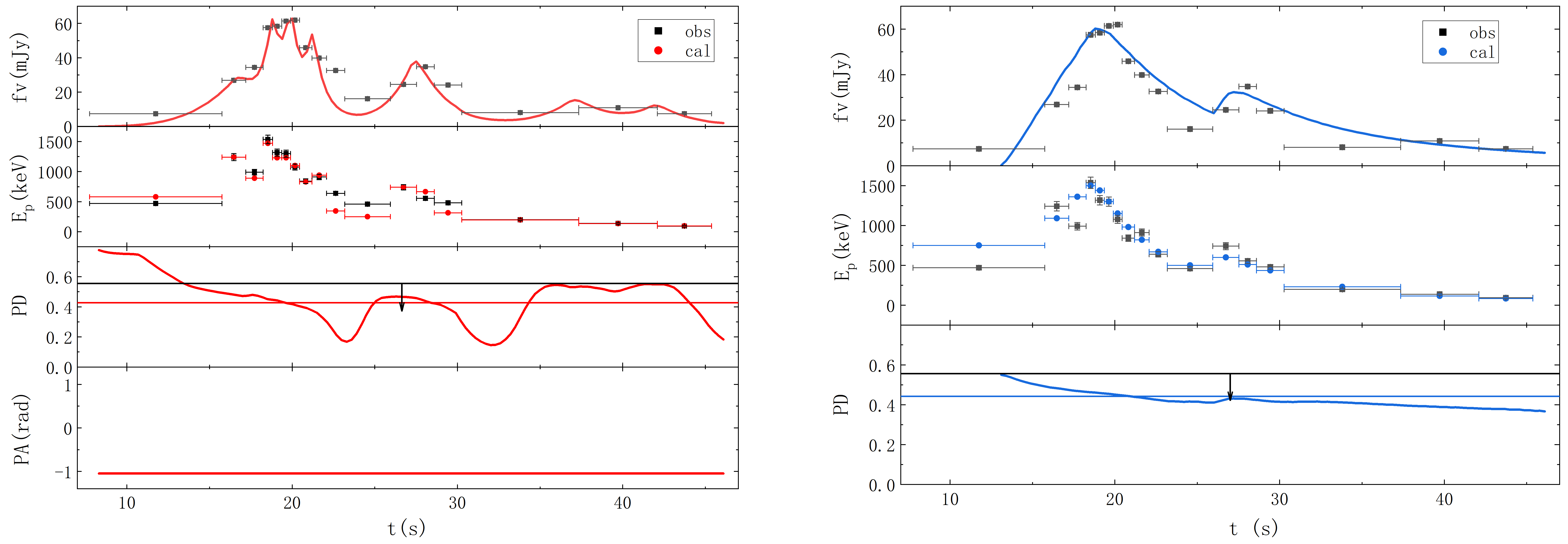}
\caption{\label{fig:171227A}Same as Figure. \ref{fig:100826A}, but for GRB 171227A.}
\end{figure*}

\begin{figure*}[h]
\centering
\includegraphics[width=0.9\linewidth]{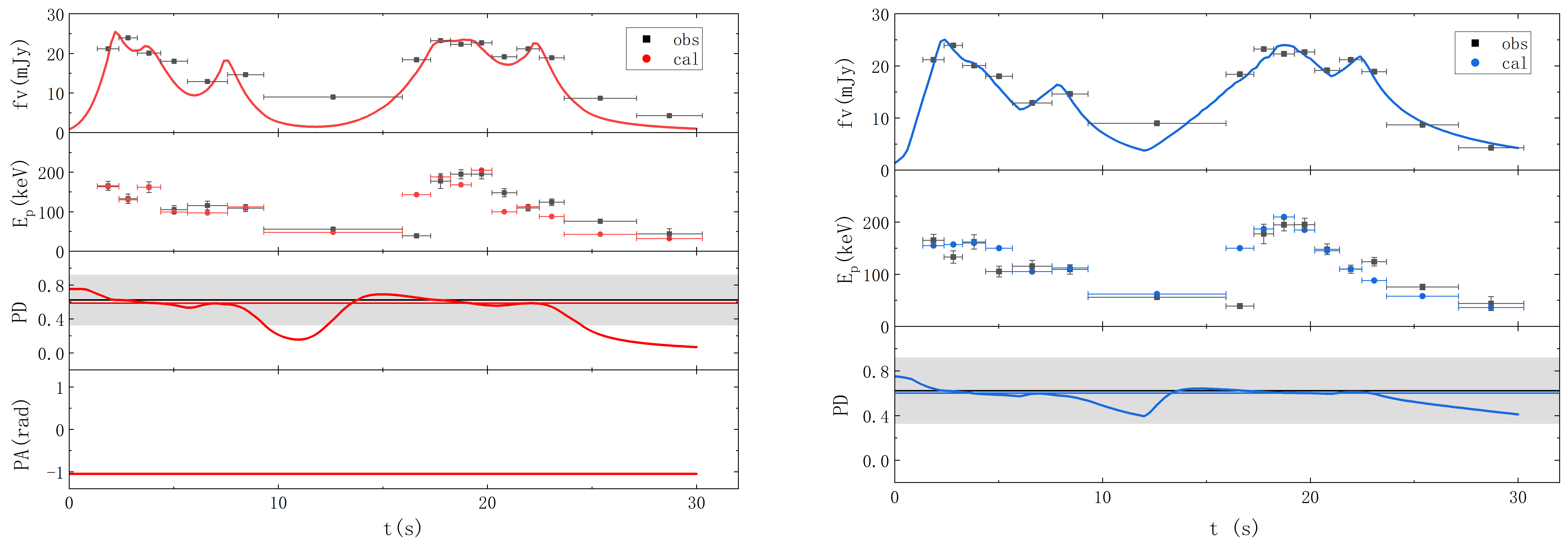}
\caption{\label{fig:180120A}Same as Figure. \ref{fig:100826A}, but for GRB 180120A.}
\end{figure*}

\begin{figure*}[h]
\centering
\includegraphics[width=0.9\linewidth]{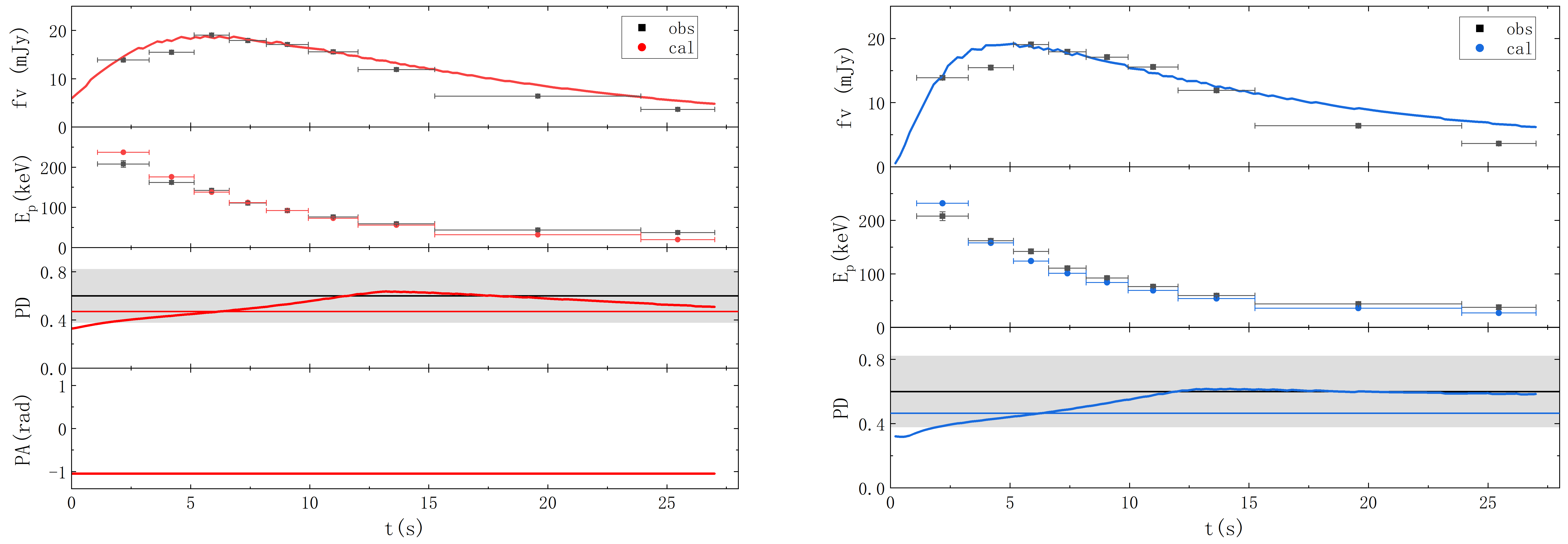}
\caption{\label{fig:180427A}Same as Figure. \ref{fig:100826A}, but for GRB 180427A.}
\end{figure*}

\begin{figure*}[h]
\centering
\includegraphics[width=0.9\linewidth]{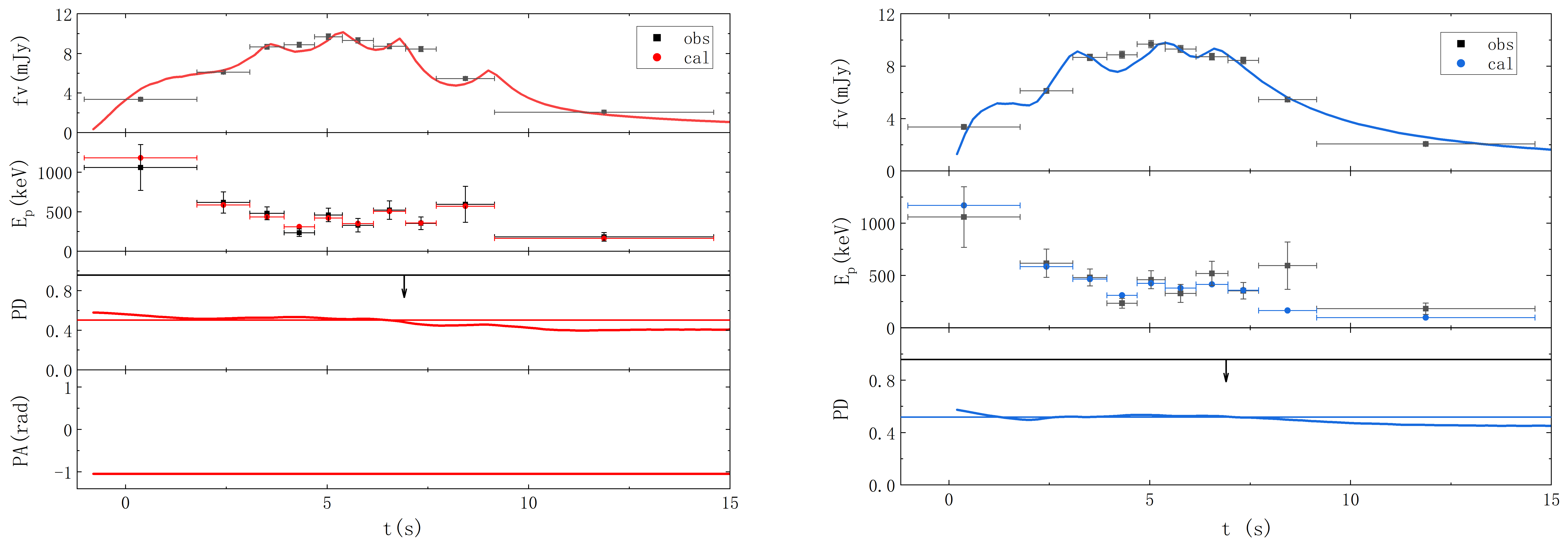}
\caption{\label{fig:180806A}Same as Figure. \ref{fig:100826A}, but for GRB 180806A.}
\end{figure*}

\begin{figure*}[h]
\centering
\includegraphics[width=0.9\linewidth]{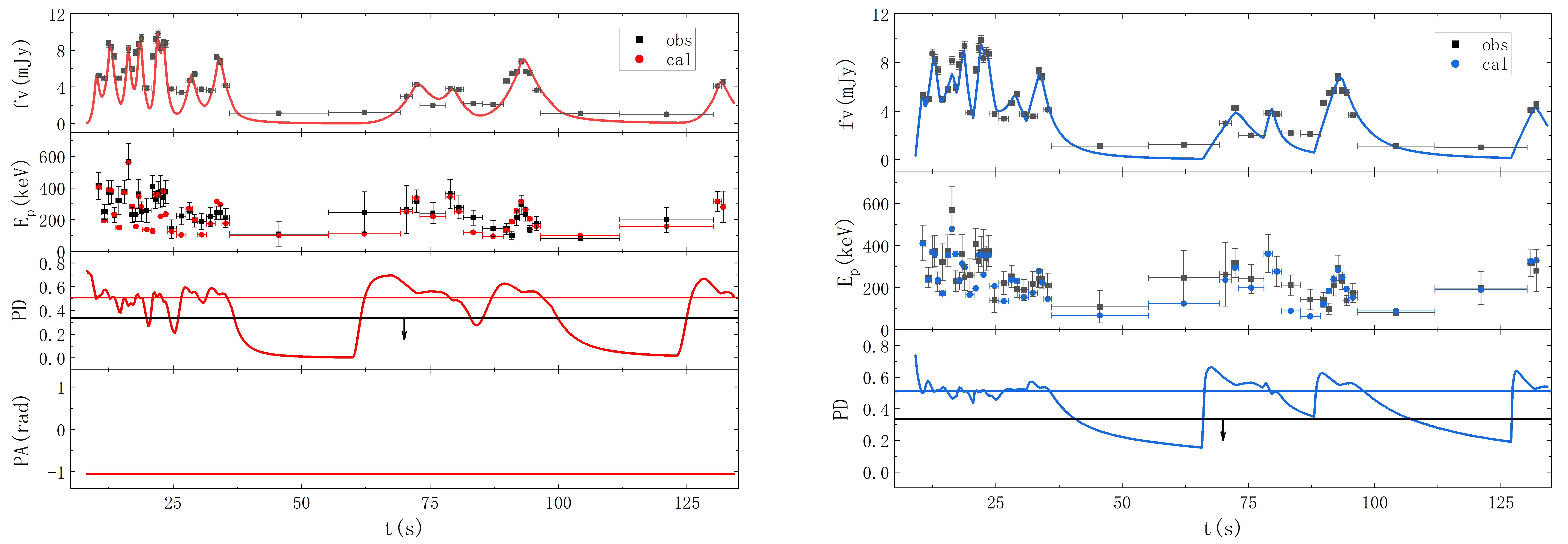}
\caption{\label{fig:180914A}Same as Figure. \ref{fig:100826A}, but for GRB 180914A.}
\end{figure*}

\begin{figure*}[h]
\centering
\includegraphics[width=0.9\linewidth]{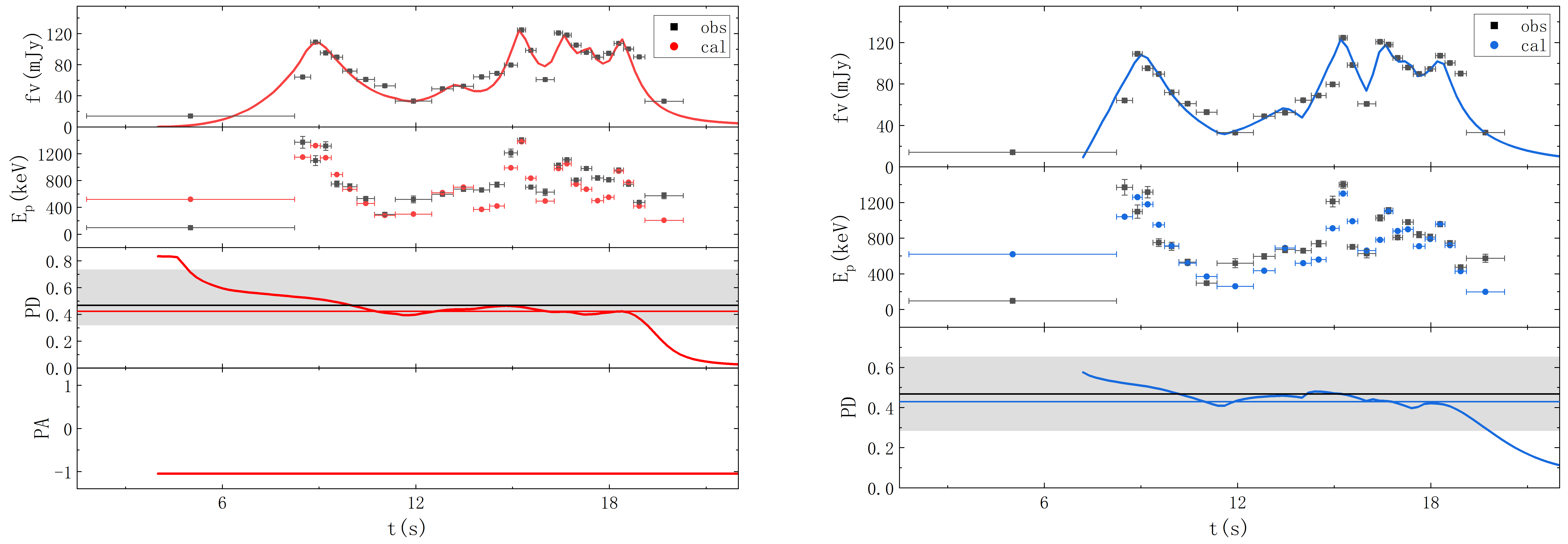}
\caption{\label{fig:190530A}Same as Figure. \ref{fig:100826A}, but for GRB 190530A.}
\end{figure*}

\begin{figure*}[h]
\centering
\includegraphics[width=0.9\linewidth]{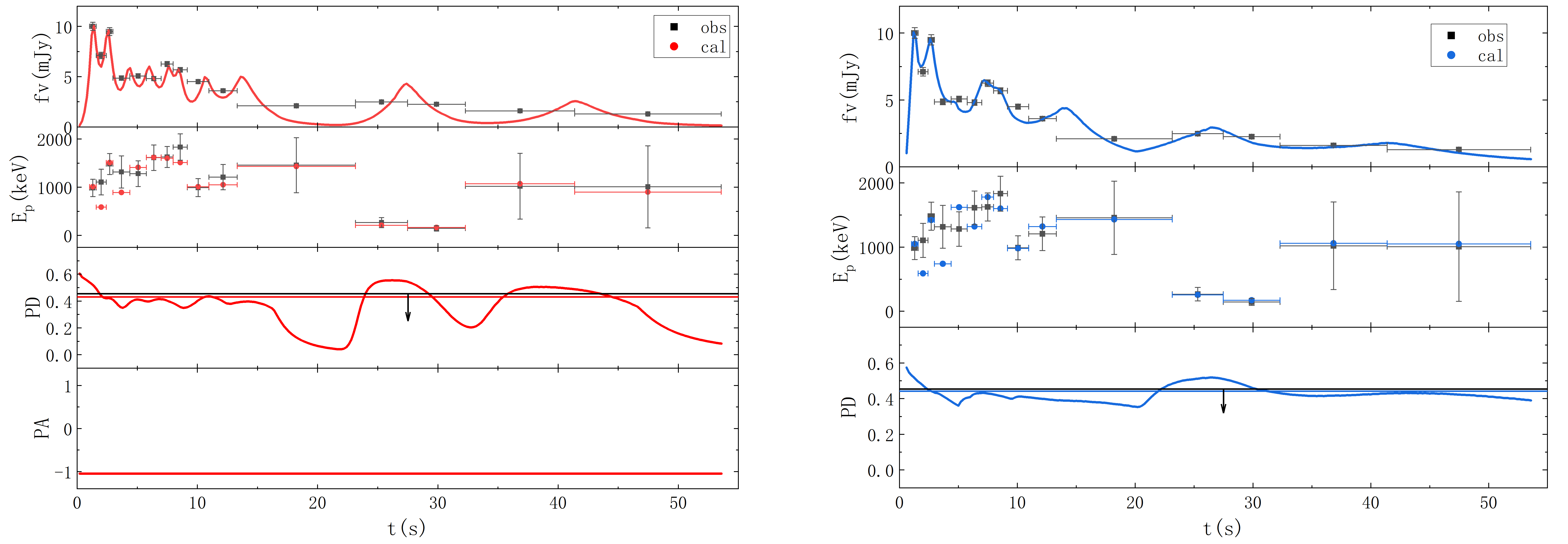}
\caption{\label{fig:200311A}Same as Figure. \ref{fig:100826A}, but for GRB 200311A.}
\end{figure*}

\begin{figure*}[h]
\centering
\includegraphics[width=0.9\linewidth]{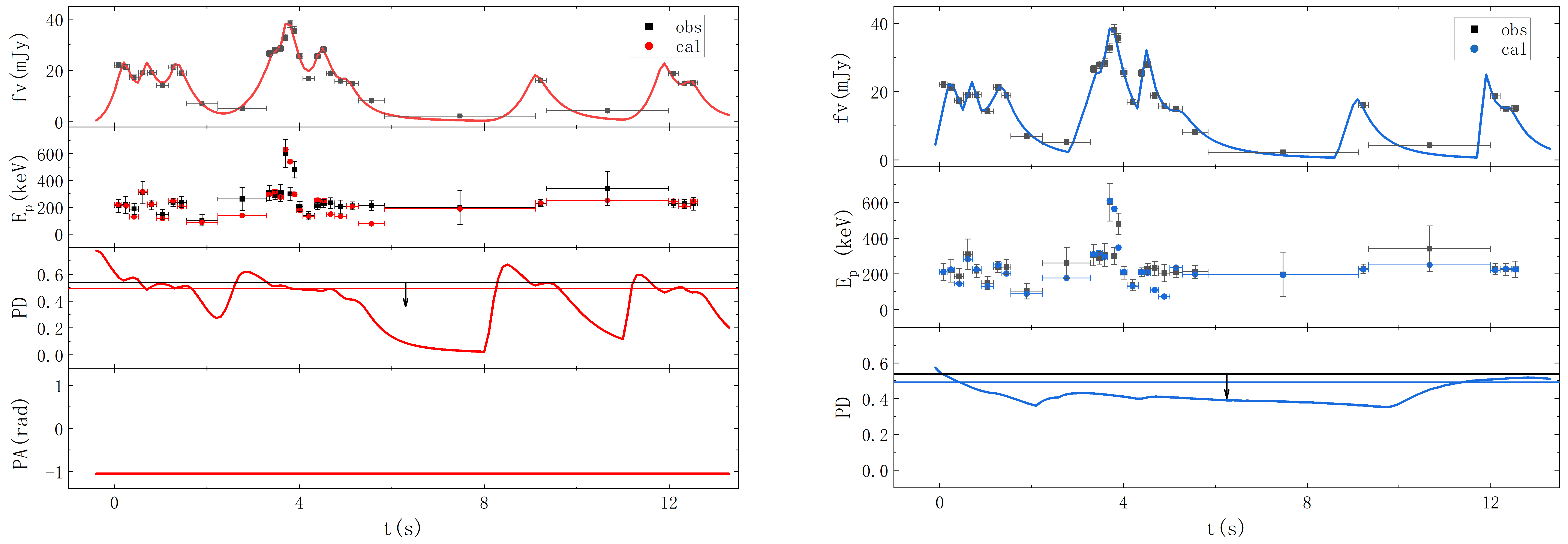}
\caption{\label{fig:200412A}Same as Figure. \ref{fig:100826A}, but for GRB of 200412A.}
\end{figure*}

Out of all 24 GRBs with multi-window observations, the low-energy spectral index of GRB 170127C is greater than 0, leading to a minus local PD $\Pi_{p,b}$. Since the local PD in an ordered magnetic field is defined to be positive, GRB 170127C is excluded in our fitting. The time-resolved multi-window fitting results of the remaining 23 GRBs are shown in Figures \ref{fig:100826A}-\ref{fig:200412A}. For GRBs 100826A, 160821A and 170114A, time-resolved polarization observations had been reported and time-resolved PAs of these bursts changed abruptly by $90^\circ$ at least once during the bursts \citep{Yonetoku_2011,Sharma_2019,Zhang_2019,Burgess_2019}. We set different orientations of the aligned magnetic fields in the radiating shells for these three GRBs to interpret the observations and the fitting details are presented in \cite{Wang_2024}. Because the reported time of GRB 100826A was not the UTC time \citep{Yonetoku_2011}, the time-resolved polarization observations is not shown in Figure. \ref{fig:100826A}. The calculation energy band for the polarizations of the GRB 160821A is same as that reported in \cite{Sharma_2019}, i.e., within (100 keV, 300 keV). Because the values of $r_0$ are larger than $3\times10^{16}$ cm, the cease radii ($r_{off}$) for GRB 160821A and GRB 180427A are set as $3\times10^{17}$ cm.

\subsection{The time-integrated and energy-integrated PD}\label{integrated}

\begin{figure}[h]
\begin{minipage}{1\linewidth}
\centering
\includegraphics[width=0.9\linewidth]{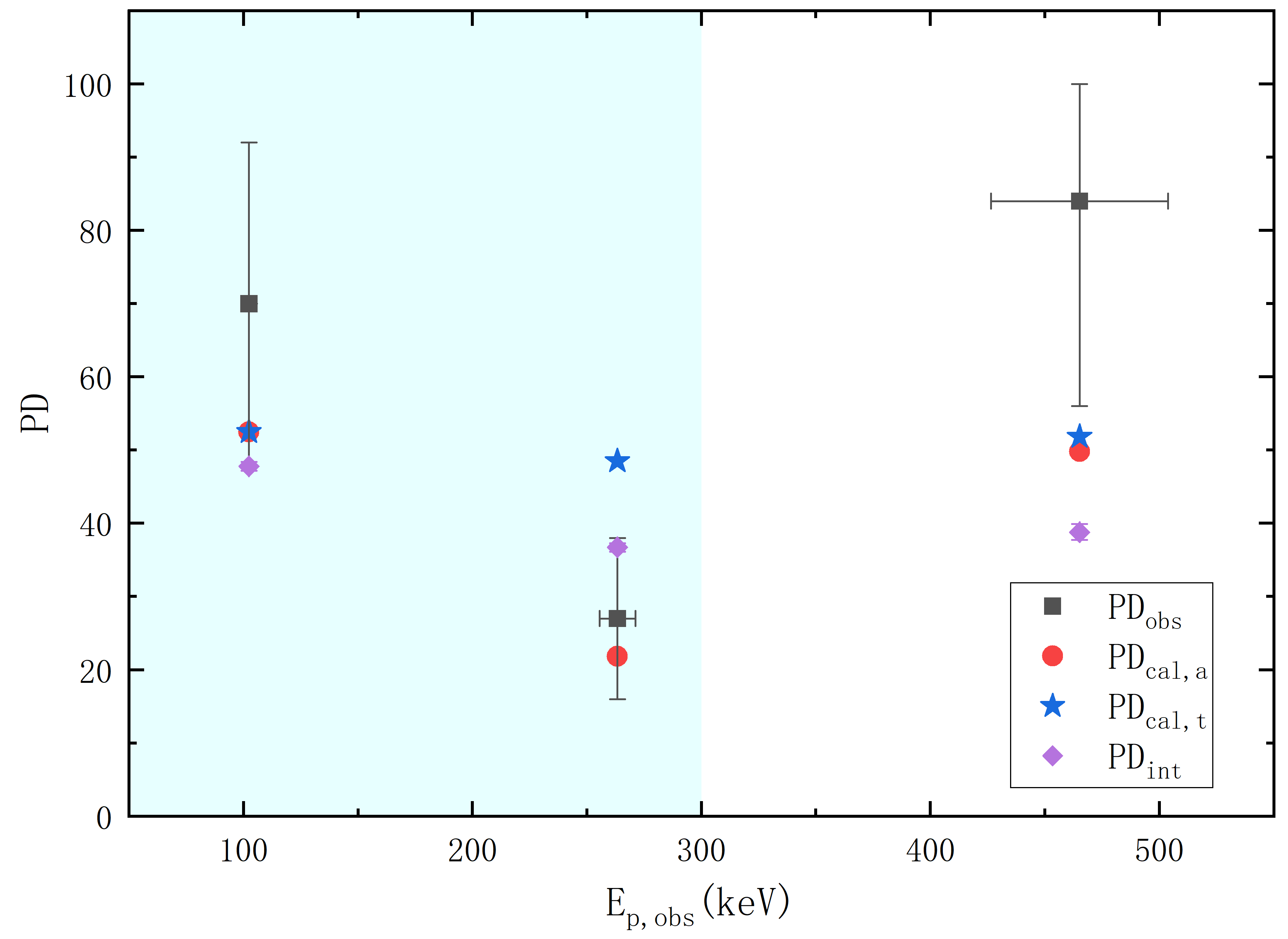}
\caption{\label{fig:GAP}Time-integrated PDs of GRBs observed by GAP. The $E_{p,obs}$ refers to the time-integrated observed value of the spectral peak energy. The black-squares, red-circles, blue-stars and purple-diamonds represents the observational values, the predicted values for the aligned-fields case, the predicted values for the toroidal-fields case and the estimated values in \cite{Guan_2023}, respectively. Polarizations here are calculated in the energy band from 70 keV to 300 keV.}
\end{minipage}
\qquad

\begin{minipage}{1\linewidth}
\centering
\includegraphics[width=0.9\linewidth]{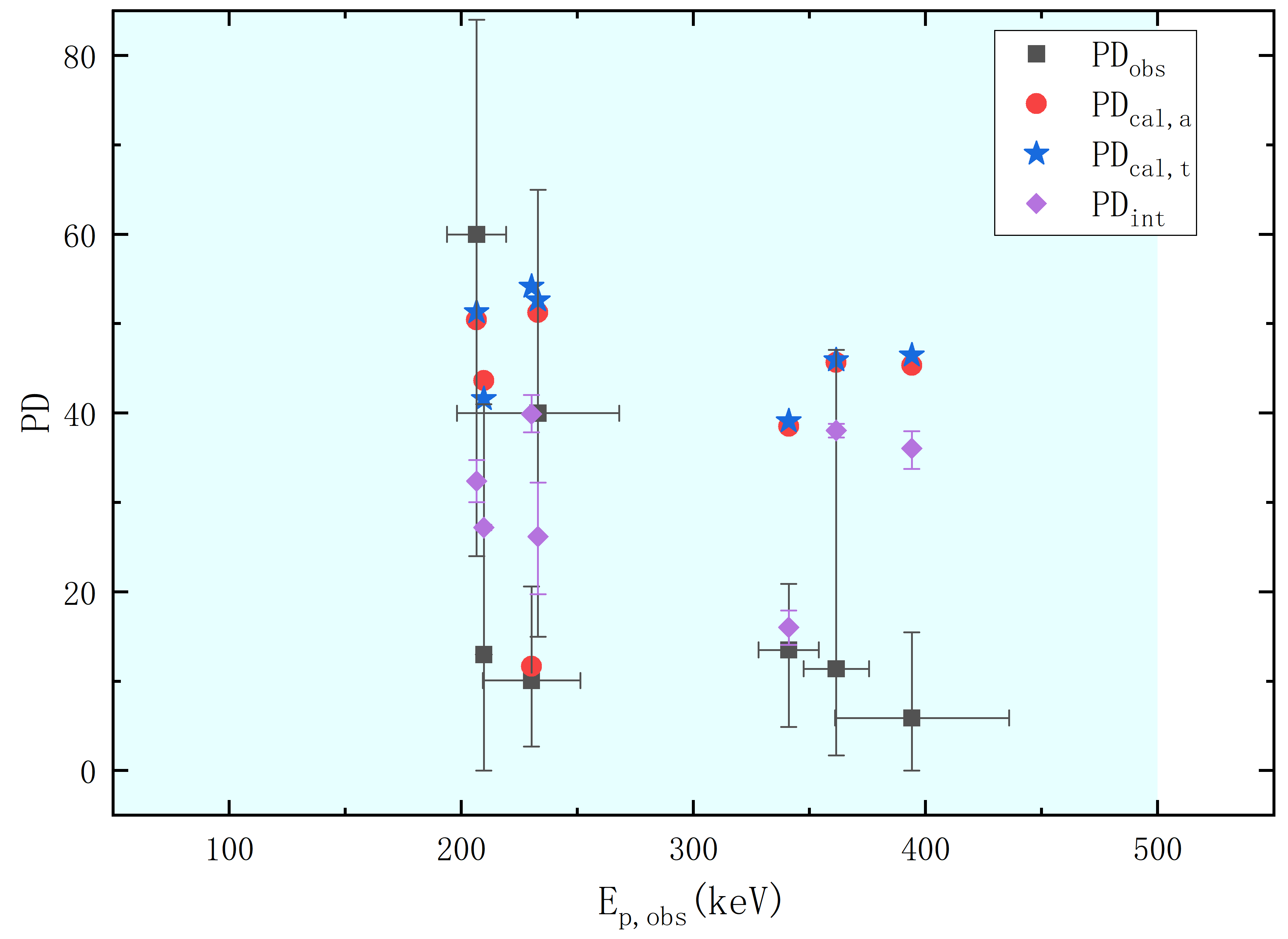}
\caption{\label{fig:POLAR}Same as Figure. \ref{fig:GAP}, but for the POLAR detected bursts in the energy range of (50 keV, 500 keV).}
\end{minipage}
\qquad

\begin{minipage}{1\linewidth}
\centering
\includegraphics[width=0.9\linewidth]{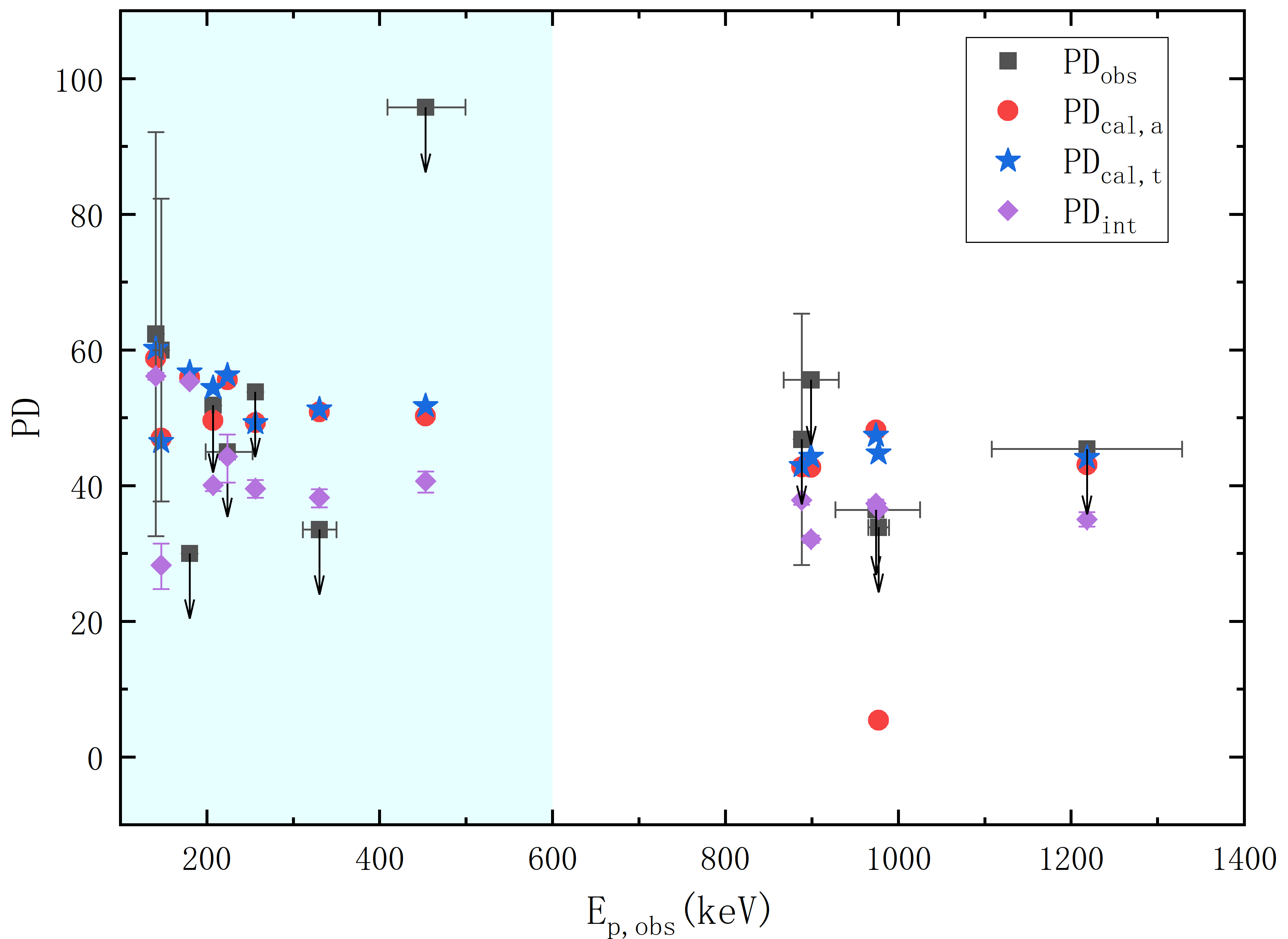}
\caption{\label{fig:AstroSat}Same as Figure. \ref{fig:GAP}, but for the AstroSat detected bursts in the energy range of (100 keV, 600 keV).}
\end{minipage}
\end{figure}

The time-resolved fitting method and results are shown in Section \ref{resolved}. With the inferred parameters (shown in Table \ref{table:parameters}) via the fitting of the multi-window observations, the time- and energy-integrated PDs ($PD_{cal,a}$ for the aligned-fields case and $PD_{cal,t}$ for the toroidal-fields case) of the 23 bursts are obtained, as shown in Figure \ref{fig:GAP} for GAP bursts, Figure \ref{fig:POLAR} for POLAR bursts and Figure \ref{fig:AstroSat} for AstroSat bursts, respectively. For comparison, the time-integrated PDs ($PD_{int}$) of these bursts calculated with the time-integrated estimation method \citep{Guan_2023} are also presented. The calculation results are summarized in Table \ref{table:integrated}. Because the orientations of the GRBs are unknown, the predicted time-integrated PAs are not compared with the corresponding observed values. For aligned-fields case, only the predicted time-resolved PAs are shown. For the toroidal-fields case, the predicted time-resolved PA would rotate abruptly by $90^\circ$ when the time-resolved PD changes its sign.

For the 20 GRBs without PA rotation observations in our sample, the directions of the aligned fields in the shells of each burst are set to be the same, leading to a constant time-resolved PA during the main radiation epoch of each burst. And both the calculated time-resolved and time-integrated PDs of the two cases (toroidal-fields case and aligned-fields case) are similar. And these predicted PDs would be the theoretical upper limits. For most of the GRBs (22 bursts out of total 23 GRBs), the predicted time-integrated PDs with two kinds of MFCs are consistent with the observations. However, there are still one burst (i.e., GRB 110721A) with observed lower limit of time-integrated PD larger than the corresponding theoretical upper limit. The PD lower limit of this bursts is given at the $1\sigma$ confidence level. Therefore, the models of the synchrotron radiation in large-scale ordered magnetic fields (both toroidal and aligned) are rejected at a confidence level of $1\sigma$ for this burst. However, the models might be still compatible with the observation in a higher confidence level. Among these 20 bursts without PA rotation observations, the observed time-integrated PDs of 9 GRBs are the upper limits and without time-resolved polarization observations. Since the predicted time- and energy-integrated PDs are the theoretical upper limits, the observed PD upper limits of these 9 bursts have no constrains on the model parameters.

The duration of GRB 170206A was divided into three time bins for polarization analysis and the observational PDs of these three time bins are $8.9^{+15.0}_{-7.3}\%$, $7.4^{+13.9}_{-5.7}\%$ and $14^{+16}_{-10}\%$, respectively \citep{Kole_2020}. The observational time-integrated PD is $13.5^{+7.4}_{-8.6}\%$, compared with the time-resolved ones, there is no reduction. Hence the time-resolved PA is inferred to be roughly a constant during the burst duration. The predicted time-resolved and time-integrated PDs of the burst are both around $\sim39\%$ for the two MFC cases, which is higher than the observed upper limit within $1\sigma$. So both the observed time-resolved and the time-integrated PDs are smaller than the corresponding predicted values. The polarization observations of the burst indicated that the magnetic field should be mixed in all its radiating shells. Similar condition exists for the first time bin of GRB 170114A, where the corresponding calculated PD for the aligned-fields case ($42.42\%$) is too high compared with the observed value ($13.21^{+6.1}_{-13.2}\%$). The polarization observations of the two bursts suggest the existence of the mixed magnetic field in at least some of the GRB radiation regions.

PAs of the three bursts (i.e., GRB 100826A, GRB 160821A and GRB 170114A) were observed to rotate with time \citep{Yonetoku_2011, Zhang_2019, Kole_2020, Sharma_2019, Chattopadhyay_2022}. To interpret such observations, the directions of the aligned magnetic fields in the adjacent shells with abrupt $90^\circ$ PA rotation observation are assumed to be differently by $90^\circ$. This assumption would lead to a reduction of the total polarized flux and hence of the final time-integrated PDs. The observed time-integrated PDs for GRB 100826A and GRB 170114A are $27_{-11}^{+11}\%$ and $10.1_{-7.4}^{+10.5}\%$, respectively \citep{Yonetoku_2011, Kole_2020}. And for GRB 160821A only an upper limit of $33.87\%$ was obtained \citep{Chattopadhyay_2022}. The predicted values for the aligned-fields case are $21.87\%$ for GRB 100826A, $5.44\%$ for GRB 160821A and $11.68\%$ for GRB 170114A, which are consistent with the observations, especially the observed best fit values. For the toroidal-fields case, the predicted time-integrated PD upper limits are $48.48\%$ for GRB 100826A, $44.76\%$ for GRB 160821A and $54.16\%$ for GRB 170114A. And the predicted time-resolved PAs are constants for these three bursts. Although the predicted time-integrated PDs are not contradict with the observations, the observed abrupt PA rotations of these three bursts can not be recovered. Therefore, only the aligned-fields case with different field orientations in the radiating shells can interpret the observed polarization properties of these three bursts. Since the aligned field are related to the magnetar central engine \citep{Spruit_2001}, as stated in a companion paper \cite{Wang_2024}, the magnetar central engines are favored for these three bursts.

\renewcommand{\arraystretch}{1.25}
\begin{table*}
\caption{The spectral parameters and the polarization properties of total 23 GRBs.\label{table:integrated}}
\begin{center}
\begin{tabular}{ccccccccc}
  \hline
  \hline
GRB & $PD_{obs}(\%)$ & $\alpha_{B}$ & $\beta_{B}$ & z & Instrument(Polarization) & $PD_{cal,a}$(\%) & $PD_{cal,t}$(\%) & $PD_{int}$(\%)\tablefootmark{a}\\
  \hline
100826A & $27_{-11}^{+11}$ & $-0.785_{-0.022}^{+0.022}$ & $-1.895_{-0.024}^{+0.024}$ & --- & GAP & 21.87 & 48.48 & $36.72_{-0.57}^{+0.57}$\\
110301A & $70_{-22}^{+22}$ & $-0.864_{-0.019}^{+0.019}$ & $-2.723_{-0.055}^{+0.055}$ & --- & GAP & 52.48 & 52.48 & $47.8_{-0.61}^{+0.59}$\\
110721A & $84_{-28}^{+16}$ & $-1.137_{-0.019}^{+0.019}$ & $-1.932_{-0.053}^{+0.053}$ & 0.382 & GAP & 49.79 & 51.85 & $38.8_{-1.07}^{+1.12}$\\
161218B & $13_{-13}^{+28}$ & $-0.597_{-0.016}^{+0.016}$ & $-3.131_{-0.206}^{+0.206}$ & --- & POLAR & 43.64 & 41.57 & $27.2_{-0.32}^{+0.34}$\\
170101B & $60_{-36}^{+24}$ & $-0.526_{-0.062}^{+0.062}$ & $-2.301_{-0.132}^{+0.132}$ & --- & POLAR & 50.43 & 51.30 & $32.42_{-2.37}^{+2.32}$\\
170114A & $10.1_{-7.4}^{+10.5}$ & $-0.928_{-0.045}^{+0.045}$ & $-2.040_{-0.075}^{+0.075}$ & --- & POLAR & 11.68 & 54.16 & $39.92_{-2.08}^{+2.1}$\\
170206A & $13.5_{-8.6}^{+7.4}$ & $-0.406_{-0.042}^{+0.042}$ & $-2.427_{-0.100}^{+0.100}$ & --- & POLAR & 38.54 & 39.12 & $16.03_{-1.95}^{+1.91}$\\
170207A & $5.9_{-5.9}^{+9.6}$ & $-0.953_{-0.025}^{+0.025}$ & $-2.433_{-0.226}^{+0.226}$ & --- & POLAR & 45.33 & 46.47 & $36.05_{-2.30}^{+1.93}$\\
170210A & $11.4_{-9.7}^{+35.7}$ & $-1.025_{-0.015}^{+0.015}$ & $-2.448_{-0.116}^{+0.116}$ & --- & POLAR & 45.66 & 45.91 & $38.06_{-0.78}^{+0.77}$\\
170305A\tablefootmark{b} & $40_{-25}^{+25}$ & $-0.242_{-0.197}^{+0.197}$ & $-1.879_{-0.083}^{+0.083}$ & --- & POLAR & 51.27 & 52.63 & $26.19_{-6.44}^{+6.04}$\\
160325A & $< 45.02$ & $-0.868_{-0.064}^{+0.064}$ & $-2.068_{-0.123}^{+0.123}$ & --- & AstroSat & 55.63 & 56.28 & $44.34_{-3.85}^{+3.24}$\\
160802A & $< 51.89$ & $-0.632_{-0.023}^{+0.023}$ & $-2.387_{-0.075}^{+0.075}$ & --- & AstroSat & 49.63 & 54.44 & $40.11_{-0.86}^{+0.27}$\\
160821A & $< 33.87$ & $-0.988_{-0.004}^{+0.004}$ & $-2.253_{-0.018}^{+0.018}$ & --- & AstroSat & 5.44 & 44.76 & $36.56_{-0.16}^{+0.16}$\\
170527A & $< 36.46$ & $-1.089_{-0.013}^{+0.013}$ & $-3.858_{-1.270}^{+1.270}$ & --- & AstroSat & 48.22 & 47.39 & $37.38_{-0.58}^{+0.57}$\\
171010A & $< 30.02$ & $-1.117_{-0.005}^{+0.005}$ & $-2.246_{-0.013}^{+0.013}$ & 0.3285 & AstroSat & 55.94 & 56.72 & $55.4_{-0.3}^{+0.22}$\\
171227A & $< 55.62$ & $-0.811_{-0.007}^{+0.007}$ & $-2.436_{-0.035}^{+0.035}$ & --- & AstroSat & 42.72 & 44.27 & $32.15_{-0.56}^{+0.56}$\\
180120A & $62.37_{-29.79}^{+29.79}$ & $-1.031_{-0.018}^{+0.018}$ & $-2.497_{-0.053}^{+0.053}$ & --- & AstroSat & 58.77 & 60.22 & $56.18_{-0.55}^{+0.47}$\\
180427A & $60.01_{-22.32}^{+22.32}$ & $-0.317_{-0.044}^{+0.044}$ & $-3.037_{-0.102}^{+0.102}$ & --- & AstroSat & 47.04 & 46.48 & $28.28_{-3.48}^{+3.2}$\\
180806A & $< 95.8$ & $-0.922_{-0.029}^{+0.029}$ & $-2.141_{-0.103}^{+0.103}$ & --- & AstroSat & 50.30 & 51.75 & $40.68_{-1.7}^{+1.41}$\\
180914A & $< 33.55$ & $-0.632_{-0.044}^{+0.044}$ & $-2.084_{-0.086}^{+0.086}$ & --- & AstroSat & 50.90 & 51.27 & $38.26_{-1.43}^{+1.26}$\\
190530A & $46.85_{-18.53}^{+18.53}$ & $-0.973_{-0.004}^{+0.004}$ & $-3.680_{-0.146}^{+0.146}$ & 0.9386 & AstroSat & 42.40 & 42.95 & $37.87_{-0.68}^{+0.1}$\\
200311A & $< 45.41$ & $-0.963_{-0.027}^{+0.027}$ & $-4.962_{-7.550}^{+7.550}$ & --- & AstroSat & 43.09 & 44.14 & $35.03_{-1.05}^{+1.09}$\\
200412A & $< 53.84$ & $-0.636_{-0.030}^{+0.030}$ & $-2.429_{-0.090}^{+0.090}$ & --- & AstroSat & 49.31 & 49.23 & $39.59_{-1.34}^{+1.29}$\\
  \hline
\end{tabular}
\end{center}
\tablefoot{
\tablefootmark{a}{$PD_{int}$ represents the results obtained in \cite{Guan_2023}}\\
\tablefootmark{b}{The variation of electron Lorentz factor is a single power law with $g=-1$ for the aligned-fields case, while it is $g=-1.5$ for the toroidal-fields case.}
}
\end{table*}

\section{Conclusions and Discussion}\label{sec:Discussion}

In this work, the time-integrated polarizations of 23 GRBs are predicted via the multi-window fitting. For the aligned-fields case, the directions of the large-scale ordered aligned magnetic fields in the shells of the individual burst are set to be the same for the bursts without observed PA rotation(s). So the calculated PDs for these bursts are the upper limits. For the large-scale ordered toroidal-fields case, both the predicted time-resolved and time-integrated PDs here are also the upper limits. Actually, for the bursts without PA rotation observations the predicted PD upper limits are similar for the two MFC cases. Although there is degeneracy between the parameters, the predicted time-integrated PDs with different sets of the parameters are roughly unchanged. Our results for the time-integrated PD are in general robust.

The 4 out of total 23 bursts have the time-resolved PA observations. For GRB 170206A, because of a comparable time-integrated PD to its time-resolved ones, the time-resolved PA is inferred to be a constant. And the observed best fit values of the time-resolved and time-integrated PDs are both smaller than the corresponding predicted upper limits. Therefore, it is inferred that mixed magnetic fields would be in the radiation regions of this burst. Among these 4 bursts, time-resolved PAs of 3 bursts show abrupt rotation(s), which would lead to a reduction of the time-integrated PDs. The observed PA rotations and the best fit values of the time-integrated PD of these 3 bursts could be recovered by the aligned-fields case with different orientations of the fields in the shells. The toroidal-fields case could not interpret such polarization observations of these 3 bursts. If the current polarization measurements for these bursts with abrupt PA rotations are roughly credible, to match with the observed time-integrated PD and the time-resolved abrupt PA rotations, the MFC in the radiation regions of these bursts would be dominated by large-scale ordered aligned-field component.

The predicted average values of the time-integrated PD here are around $44\%$ for the aligned-fields case and around $49\%$ for the toroidal-fields case, both of which are higher than the results of $37.4\%$ in \cite{Guan_2023}. Compared to the constant time-resolved PAs predicted in the toroidal-fields case, the predicted PA rotations of the aligned-fields case lead to a smaller average value of the time-integrated PD. The results of the time-integrated estimation method used in \cite{Guan_2023, Toma_2009} are relatively smaller than the results here. Because the predicted time-resolved PA is a constant during the main radiation epoch of each burst for the toroidal-fields case, the result here is comparable to that of $50\%$ in \cite{SL2024}. Compared with the typical or fiducial parameters used in \cite{SL2024}, the model parameters are inferred from the multi-window fitting here. Although the time-integrated PDs are both obtained via the time-resolved ones here and in \cite{SL2024}, the results here should be more accurate, especially for the bursts with PA rotations.  

For most of the bursts (22 out of total 23 bursts) in our sample, the predicted time- and energy-integrated PDs could match with the corresponding observations. There is no time-resolved PA observations or the time-resolved PA is indicated to be a constant for 19 bursts among these 22 bursts. Both the aligned-fields case and the toroidal-fields case could interpret the observational data of these 19 bursts equally well. However, the calculated upper limits for the time-integrated PD of one burst in our sample are lower than the corresponding observed lower limit of $1\sigma$ confidence level. Since the observational errors of the polarizations are large and the confidence level is relatively low, whether or not the models of the synchrotron emission in ordered magnetic fields are rejected for the this burst is not sure. Finally, the synchrotron radiation in ordered magnetic field(s) could interpret the multi-window observations of most bursts in our sample. Future accurate polarization measurements will put strict constrains on the MFC in the radiation region and the origin of the GRBs.

\begin{acknowledgements}

{We are grateful for the GRB data of Fermi/GBM, GAP, POLAR and AstroSat. This work is supported by the National Natural Science Foundation of China (grant Nos. 12473040, 11903014, 12321003, 12393810, 11833003, 11903017 and 12065017), the National Key Research and Development Program of China (grant no. 2017YFA0402600), the National SKA Program of China No. 2020SKA0120300, International Partnership Program of Chinese Academy of Sciences for Grand Challenges (114332KYSB20210018), Jiangxi Provincial Natural Science Foundation under grant 20224ACB211001 and the Major Science and Technology Project of Qinghai Province (2019-ZJ-A10). M.X.L also would like to appreciate the financial support from Jilin University.}

\end{acknowledgements}

\bibliographystyle{aa}
\bibliography{ms_arXiv}

\begin{appendix}

\section{The parameters used in the time-resolved multi-window fitting}\label{appC}

\onecolumn
\begin{landscape}
\renewcommand{\arraystretch}{1.25}
\begin{longtable}{c|ccccccc|cccccc}
\caption{The parameters used in fitting. \label{table:parameters}}\\
\hline
\hline
GRB & Shell & $t_0$(s) & $\gamma_{ch}^{0}/\gamma_{ch}^{m}$ & $r_0$(cm) & $r_m$(cm) & $R_{inj}$($s^{-1}$) & $\delta$(rad) & Shell & $t_0$ & $\Gamma_{0}$ & $\gamma_{ch}^{0}/\gamma_{ch}^{m}$ & $r_m$(cm) & $R_{inj}$($s^{-1}$) \\
\hline
\endfirsthead
\caption{continued.} \\
\hline
\hline
GRB & Shell & $t_0$(s) & $\gamma_{ch}^{0}/\gamma_{ch}^{m}$ & $r_0$(cm) & $r_m$(cm) & $R_{inj}$($s^{-1}$) & $\delta$(rad) & Shell & $t_0$ & $\Gamma_{0}$ & $\gamma_{ch}^{0}/\gamma_{ch}^{m}$ & $r_m$(cm) & $R_{inj}$($s^{-1}$) \\
\hline
\endhead
\hline
\endfoot
\hline
\endlastfoot
 100826A & 1 & 0 & $7.2\times 10^4$ & $1.45\times 10^{16}$ & $1.4\times 10^{16}$ & $1.04\times 10^{49}$ & $-\pi/10$
& 1 & 8 & $200$ & $3\times 10^5$ & $1.45\times 10^{16}$ & $2.4\times 10^{50}$\\
& 2 & 25 & $7.2\times 10^4$ & $3.2\times 10^{15}$ & $5\times 10^{15}$ & $1.72\times 10^{49}$ & $-\pi/10$
& 2 & 60 & $200$ & $2\times 10^5$ & $1\times 10^{16}$ & $6.4\times 10^{49}$\\
 & 3 & 60 & $6.3\times 10^4$ & $7.3\times 10^{15}$ & $1\times 10^{16}$ & $1\times 10^{49}$ & $-\pi/10+\pi/2$
& 3 & 68 & $200$ & $1.8\times 10^5$ & $1.3\times 10^{16}$ & $1.28\times 10^{50}$\\
 & 4 & 70 & $5.7\times 10^4$ & $5.5\times 10^{15}$ & $1\times 10^{16}$ & $1.8\times 10^{49}$ & $-\pi/10+\pi/2$ &&&&&&\\
  \hline
110301A & 1 & -1 & $7.9\times 10^4$ & $1\times 10^{15}$ & $2.8\times 10^{15}$ & $8\times 10^{50}$ & $\pi/6$
& 1 & -1 & $200$ & $7.5\times 10^4$ & $1.9\times 10^{15}$ & $1.52\times 10^{51}$\\
 & 2 & 0 & $7.1\times 10^4$ & $1\times 10^{15}$ & $3\times 10^{15}$ & $1.6\times 10^{51}$ & $\pi/6$
& 2 & 0.6 & $200$ & $5.8\times 10^4$ & $1.1\times 10^{15}$ & $1.92\times 10^{51}$\\
 & 3 & 1 & $7.5\times 10^4$ & $1\times 10^{15}$ & $2.3\times 10^{15}$ & $2.36\times 10^{51}$ & $\pi/6$
& 3 & 1.6 & $200$ & $6.5\times 10^4$ & $1.1\times 10^{15}$ & $3.76\times 10^{51}$\\
& 4 & 2.8 & $6\times 10^4$ & $1\times 10^{15}$ & $1.4\times 10^{15}$ & $3.2\times 10^{51}$ & $\pi/6$
& 4 & 3.4 & $200$ & $5\times 10^4$ & $7\times 10^{14}$ & $4.4\times 10^{51}$\\
  \hline
110721A & 1 & 0 & $4.6\times 10^4$ & $3.1\times 10^{15}$ & --- & $2\times 10^{48}$ & $\pi/6$
& 1 & 0.2 & $85$ & $9.5\times 10^4$ & --- & $1.04\times 10^{49}$\\
 & 2 & 1.4 & $1.3\times 10^4$ & $6.7\times 10^{15}$ & --- & $2.6\times 10^{48}$ & $\pi/6$
& 2 & 1.8 & $85$ & $3.8\times 10^4$ & --- & $2.4\times 10^{49}$\\
 & 3 & 5 & $1.28\times 10^5$ & $4.6\times 10^{15}$ & $1\times 10^{16}$ & $4\times 10^{47}$ & $\pi/6$ &&&&&&\\
  \hline
161218B & 1 & 1.6 & $1.5\times 10^4$ & $2\times 10^{15}$ & --- & $3.48\times 10^{51}$ & $\pi/6$
& 1 & 1.4 & $150$ & $5.6\times 10^4$ & $4.5\times 10^{14}$ & $4\times 10^{51}$\\
& 2 & 4 & $9.3\times 10^4$ & $1\times 10^{15}$ & $2.6\times 10^{15}$ & $6.8\times 10^{50}$ & $\pi/6$
& 2 & 4.5 & $150$ & $7.5\times 10^4$ & $8.5\times 10^{14}$ & $1.6\times 10^{51}$\\
& 3 & 6 & $1.15\times 10^5$ & $1.2\times 10^{15}$ & $5\times 10^{15}$ & $3.36\times 10^{50}$ & $\pi/6$
& 3 & 7 & $150$ & $9\times 10^4$ & $9.6\times 10^{14}$ & $5.6\times 10^{50}$\\
& 4 & 8.5 & $1.08\times 10^5$ & $1.2\times 10^{15}$ & $4.1\times 10^{15}$ & $3.04\times 10^{50}$ & $\pi/6$
& 4 & 9 & $150$ & $1\times 10^5$ & $1.1\times 10^{15}$ & $3.4\times 10^{50}$\\
& 5 & 10 & $1.1\times 10^5$ & $1.2\times 10^{15}$ & $4\times 10^{15}$ & $2.08\times 10^{50}$ & $\pi/6$
& 5 & 10 & $150$ & $9.6\times 10^4$ & $1.3\times 10^{15}$ & $3.6\times 10^{50}$\\
& 6 & 11 & $1\times 10^5$ & $2\times 10^{15}$ & $5\times 10^{15}$ & $1.04\times 10^{50}$ & $\pi/6$
& 6 & 12 & $150$ & $1.2\times 10^5$ & $1.5\times 10^{15}$ & $2.52\times 10^{50}$\\
& 7 & 15 & $8.3\times 10^4$ & $2\times 10^{15}$ & $5\times 10^{15}$ & $1.52\times 10^{50}$ & $\pi/6$
& 7 & 16 & $150$ & $1\times 10^5$ & $1.3\times 10^{15}$ & $3.2\times 10^{50}$\\
& 8 & 19 & $6\times 10^3$ & $1.8\times 10^{16}$ & --- & $2.8\times 10^{49}$ & $\pi/6$
& 8 & 19 & $150$ & $1.5\times 10^5$ & $1.2\times 10^{15}$ & $9.2\times 10^{49}$\\
&&&&&&&& 9 & 21.5 & $150$ & $9.2\times 10^4$ & $2\times 10^{15}$ & $7.2\times 10^{50}$\\
  \hline
170101B & 1 & 0 & $6.5\times 10^3$ & $1\times 10^{16}$ & --- & $2.12\times 10^{49}$ & $\pi/6$
& 1 & 0.2 & $60$ & $3\times 10^4$ & --- & $3.36\times 10^{50}$\\
 & 2 & 3 & $1.06\times 10^5$ & $2.3\times 10^{15}$ & $1\times 10^{16}$ & $2.72\times 10^{49}$ & $\pi/6$
& 2 & 3.5 & $60$ & $9.2\times 10^4$ & $5.5\times 10^{14}$ & $8.4\times 10^{49}$\\
& 3 & 6 & $9.7\times 10^4$ & $2.2\times 10^{15}$ & $1\times 10^{16}$ & $3.6\times 10^{49}$ & $\pi/6$
& 3 & 6 & $60$ & $8.2\times 10^4$ & $5.9\times 10^{14}$ & $9.6\times 10^{49}$\\
  \hline
170114A & 1 & 0 & $1.45\times 10^4$ & $5.2\times 10^{15}$ & --- & $2.6\times 10^{49}$ & $-\pi/6$
& 1 & 0 & 70 & $4.6\times 10^4$ & --- & $1.8\times 10^{50}$ \\
& 2 & 0.5 & $4.9\times 10^4$ & $2.2\times 10^{15}$ & $1.1\times 10^{15}$ & $8\times 10^{49}$ & $-\pi/6-\pi/2$
& 2 & 1 & 100 & $6.3\times 10^4$ & $4.5\times 10^{14}$ & $3.48\times 10^{50}$\\
& 3 & 1.2 & $4.85\times 10^4$ & $2.4\times 10^{15}$ & $1.2\times 10^{15}$ & $2.84\times 10^{49}$ & $-\pi/6-\pi/2$
& 3 & 6 & 100 & $8.5\times 10^4$ & $1\times 10^{15}$ & $4\times 10^{49}$\\
& 4 & 5 & $6.8\times 10^3$ & $8\times 10^{15}$ & --- & $1.32\times 10^{49}$ & $-\pi/6$ &&&&&&\\
  \hline
170206A & 1 & 0 & $6.8\times 10^4$ & $7\times 10^{14}$ & $4.4\times 10^{14}$ & $1.08\times 10^{51}$ & $\pi/6$
& 1 & 0.2 & $200$ & $5.1\times 10^4$ & $3.5\times 10^{14}$ & $1.12\times 10^{51}$\\
 & 2 & 0.6 & $7.2\times 10^4$ & $2.5\times 10^{14}$ & $2\times 10^{14}$ & $3.8\times 10^{51}$ & $\pi/6$
& 2 & 0.6 & $200$ & $3.6\times 10^4$ & $2\times 10^{14}$ & $1.32\times 10^{51}$\\
 & 3 & 0.8 & $5.3\times 10^4$ & $6.2\times 10^{14}$ & $4\times 10^{14}$ & $1.4\times 10^{51}$ & $\pi/6$
& 3 & 0.9 & $200$ & $4.1\times 10^4$ & $3.7\times 10^{14}$ & $8.8\times 10^{50}$\\
  \hline
170207A & 1 & 0 & $1\times 10^5$ & $1\times 10^{15}$ & $6.5\times 10^{15}$ & $3.4\times 10^{50}$ & $\pi/6$
& 1 & 1 & $150$ & $6.5\times 10^4$ & $1.1\times 10^{15}$ & $6\times 10^{50}$\\
 & 2 & 3.8 & $1.1\times 10^5$ & $1\times 10^{15}$ & $1.2\times 10^{15}$ & $1.28\times 10^{50}$ & $\pi/6$
& 2 & 4.2 & $150$ & $9.4\times 10^4$ & $5.2\times 10^{14}$ & $2.12\times 10^{50}$\\
 & 3 & 4 & $2\times 10^5$ & $1\times 10^{15}$ & $3.7\times 10^{15}$ & $5.2\times 10^{49}$ & $\pi/6$
& 3 & 5.2 & $150$ & $1.35\times 10^5$ & $5\times 10^{14}$ & $5.6\times 10^{49}$\\
& 4 & 6 & $1.5\times 10^5$ & $1\times 10^{15}$ & $1.8\times 10^{15}$ & $4.8\times 10^{49}$ & $\pi/6$
& 4 & 6 & $150$ & $1.4\times 10^5$ & $1.1\times 10^{15}$ & $6.4\times 10^{49}$\\
& 5 & 14 & $2.4\times 10^5$ & $1.3\times 10^{15}$ & $1\times 10^{16}$ & $8.8\times 10^{49}$ & $\pi/6$
& 5 & 16 & $150$ & $1.45\times 10^5$ & $9\times 10^{14}$ & $2.04\times 10^{50}$\\
& 6 & 17.5 & $1.75\times 10^5$ & $1\times 10^{15}$ & $1.4\times 10^{15}$ & $6.4\times 10^{49}$ & $\pi/6$
& 6 & 18 & $150$ & $1.3\times 10^5$ & $5\times 10^{14}$ & $1.2\times 10^{50}$\\
& 7 & 19 & $1.65\times 10^5$ & $1\times 10^{15}$ & $2.8\times 10^{15}$ & $1.48\times 10^{50}$ & $\pi/6$
& 7 & 20 & $150$ & $1.1\times 10^5$ & $5.8\times 10^{14}$ & $2.52\times 10^{50}$\\
& 8 & 20 & $1.5\times 10^5$ & $1\times 10^{15}$ & $2\times 10^{15}$ & $7.2\times 10^{49}$ & $\pi/6$
& 8 & 20.7 & $150$ & $1.1\times 10^5$ & $5.5\times 10^{14}$ & $1.32\times 10^{50}$\\
  \hline
170210A& 1 & 34 & $2\times 10^5$ & $1.7\times 10^{15}$ & $1\times 10^{16}$ & $4.8\times 10^{49}$ &$\pi/6$
& 1 & 35 & $150$ & $1.9\times 10^5$ & $1.9\times 10^{15}$ & $1.36\times 10^{50}$\\
  & 2 & 39 & $1.85\times 10^5$ & $1\times 10^{15}$ & $3.5\times 10^{15}$ & $8.8\times 10^{49}$ & $\pi/6$
& 2 & 39 & $150$ & $1.6\times 10^5$ & $1.1\times 10^{15}$ & $1\times 10^{50}$\\
  & 3 & 40 & $2.1\times 10^5$ & $1\times 10^{15}$ & $5\times 10^{15}$ & $1.08\times 10^{50}$ & $\pi/6$
& 3 & 40 & $150$ & $1.8\times 10^5$ & $1.4\times 10^{15}$ & $1.6\times 10^{50}$\\
  & 4 & 41 & $2.55\times 10^5$ & $1\times 10^{15}$ & $7.2\times 10^{15}$ & $6.8\times 10^{49}$ & $\pi/6$
& 4 & 46 & $150$ & $2.9\times 10^5$ & $3.2\times 10^{15}$ & $5.6\times 10^{49}$\\
  & 5 & 45 & $2\times 10^5$ & $3.2\times 10^{15}$ & $1\times 10^{16}$ & $1.12\times 10^{49}$ & $\pi/6$
& 5 & 53 & $150$ & $2.6\times 10^5$ & $3.4\times 10^{15}$ & $1.08\times 10^{50}$\\
  & 6 & 48 & $1.4\times 10^5$ & $5.7\times 10^{15}$ & $1.2\times 10^{16}$ & $1.36\times 10^{49}$ & $\pi/6$
& 6 & 65 & $150$ & $1.8\times 10^5$ & $1.7\times 10^{15}$ & $1.96\times 10^{50}$\\
  & 7 & 60 & $1.3\times 10^5$ & $4.5\times 10^{15}$ & $1\times 10^{16}$ & $3.12\times 10^{49}$ & $\pi/6$ &&&&&&\\
  \hline
170305A\tablefootmark{a}& 1 & -0.02 & $1.6\times 10^4$ & $3\times 10^{14}$ & --- & $5.2\times 10^{51}$ &$\pi/6$
& 1 & -0.02 & $180$ & $8.6\times 10^2$ & --- & $2.6\times 10^{51}$\\
  \hline
160325A & 1 & 1 & $1.1\times 10^5$ & $1\times 10^{15}$ & $6.5\times 10^{15}$ & $8.4\times 10^{49}$ & $\pi/6$
& 1 & 2 & $150$ & $7.5\times 10^4$ & $9.5\times 10^{14}$ & $1.24\times 10^{50}$\\
& 2 & 3.5 & $1\times 10^5$ & $1\times 10^{15}$ & $4\times 10^{15}$ & $8.8\times 10^{49}$ & $\pi/6$
& 2 & 4 & $150$ & $7.5\times 10^4$ & $9\times 10^{14}$ & $1\times 10^{50}$\\
 & 3 & 5 & $1.05\times 10^5$ & $1.6\times 10^{15}$ & $1\times 10^{16}$ & $6\times 10^{49}$ & $\pi/6$
& 3 & 7 & $150$ & $8\times 10^4$ & $1.3\times 10^{15}$ & $1.6\times 10^{50}$\\
& 4 & 9 & $1.3\times 10^5$ & $1\times 10^{15}$ & $4.4\times 10^{15}$ & $3.48\times 10^{49}$ & $\pi/6$
& 4 & 9 & $150$ & $1.2\times 10^5$ & $1.45\times 10^{15}$ & $4.8\times 10^{49}$\\
& 5 & 10 & $1.4\times 10^5$ & $1\times 10^{15}$ & $1.3\times 10^{16}$ & $6\times 10^{49}$ & $\pi/6$
& 5 & 12 & $150$ & $7.2\times 10^4$ & $7.6\times 10^{14}$ & $6\times 10^{49}$\\
& 6 & 38 & $5\times 10^4$ & $2.9\times 10^{15}$ & $1\times 10^{16}$ & $5.2\times 10^{49}$ & $\pi/6$
& 6 & 39 & $150$ & $7.6\times 10^4$ & $3.6\times 10^{15}$ & $2.64\times 10^{50}$\\
  \hline
160802A & 1 & 0 & $1.6\times 10^4$ & $4\times 10^{15}$ & --- & $4\times 10^{50}$ & $\pi/6$
& 1 & 0.2 & $80$ & $3.8\times 10^4$ & --- & $2.68\times 10^{51}$\\
 & 2 & 16 & $2.7\times 10^4$ & $1\times 10^{15}$ & $4\times 10^{14}$ & $2.92\times 10^{51}$ & $\pi/6$
& 2 & 16.2 & $80$ & $2.3\times 10^4$ & $1.5\times 10^{14}$ & $1.12\times 10^{52}$\\
  \hline
160821A & 1 & 117 & $8.8\times 10^3$ & $5.5\times 10^{16}$ & --- & $2\times 10^{48}$ & $\pi/6$
&1 & 118 & 33 & $1.4\times 10^5$ & --- & $1.6\times 10^{50}$\\
 & 2 & 122 & $8.1\times 10^4$ & $1.5\times 10^{16}$ & $1.9\times 10^{15}$ & $7.2\times 10^{48}$ & $\pi/6+\pi/2$
&2 & 124 & 100 & $3.5\times 10^5$ & $2\times 10^{15}$ & $1.48\times 10^{50}$\\
 & 3 & 130 & $7.1\times 10^4$ & $1.35\times 10^{16}$ & $1\times 10^{15}$ & $1.32\times 10^{49}$ & $\pi/6+\pi/2$
&3 & 131 & 100 & $3.4\times 10^5$ & $1.5\times 10^{15}$ & $1.52\times 10^{50}$\\
 & 4 & 134 & $9.5\times 10^4$ & $7.8\times 10^{15}$ & $1\times 10^{15}$ & $1.2\times 10^{49}$ & $\pi/6+\pi/2$
&4 & 134 & 100 & $2.5\times 10^5$ & $1.2\times 10^{15}$ & $1.6\times 10^{50}$\\
 & 5 & 140 & $6.5\times 10^4$ & $1.1\times 10^{16}$ & $1.1\times 10^{15}$ & $8.4\times 10^{48}$ & $\pi/6$
&5 & 143 & 100 & $2.5\times 10^5$ & $7.4\times 10^{14}$ & $4\times 10^{49}$\\  
\hline
170527A & 1 & 0 & $2.2\times 10^5$ & $1\times 10^{15}$ & $2.7\times 10^{15}$ & $7.6\times 10^{49}$ & $\pi/6$
& 1 & 0.5 & $150$ & $1.6\times 10^5$ & $8\times 10^{14}$ & $1.64\times 10^{50}$\\
 & 2 & 2 & $4.1\times 10^5$ & $1\times 10^{15}$ & $8.2\times 10^{15}$ & $3.88\times 10^{49}$ & $\pi/6$
& 2 & 3 & $150$ & $2.5\times 10^5$ & $1.1\times 10^{15}$ & $7.6\times 10^{49}$\\
 & 3 & 5.5 & $3.4\times 10^5$ & $1\times 10^{15}$ & $4.3\times 10^{15}$ & $2.6\times 10^{49}$ & $\pi/6$
& 3 & 5.8 & $150$ & $2.7\times 10^5$ & $1.05\times 10^{15}$ & $3.4\times 10^{49}$\\
 & 4 & 9 & $2.6\times 10^5$ & $2.6\times 10^{15}$ & $9.5\times 10^{15}$ & $1.4\times 10^{49}$ & $\pi/6$
& 4 & 10 & $150$ & $3.2\times 10^5$ & $2.8\times 10^{15}$ & $6.4\times 10^{49}$\\
 & 5 & 18 & $9\times 10^3$ & $2.7\times 10^{16}$ & --- & $3.6\times 10^{48}$ & $\pi/6$
& 5 & 18.5 & $150$ & $3.1\times 10^5$ & $2.3\times 10^{15}$ & $3.32\times 10^{49}$\\
& 6 & 27 & $1.5\times 10^5$ & $4.3\times 10^{15}$ & $1\times 10^{16}$ & $8.8\times 10^{48}$ & $\pi/6$
& 6 & 23 & $150$ & $2.3\times 10^5$ & $2.1\times 10^{15}$ & $4.4\times 10^{49}$\\
&&&&&&&& 7 & 27 & $150$ & $2.3\times 10^5$ & $1.8\times 10^{15}$ & $3\times 10^{49}$\\
&&&&&&&& 8 & 31 & $150$ & $1.6\times 10^5$ & $2.5\times 10^{15}$ & $1\times 10^{50}$\\
  \hline
171010A & 1 & 7 & $7.3\times 10^4$ & $1.2\times 10^{16}$ & $2\times 10^{16}$ & $4.8\times 10^{48}$ & $\pi/6$
& 1 & 12 & $150$ & $2.2\times 10^5$ & $8.5\times 10^{15}$ & $1\times 10^{50}$\\
 & 2 & 21 & $8.4\times 10^4$ & $3.2\times 10^{15}$ & $1\times 10^{16}$ & $1.72\times 10^{49}$ & $\pi/6$
& 2 & 22 & $150$ & $1.15\times 10^5$ & $3.5\times 10^{15}$ & $7.2\times 10^{49}$\\
 & 3 & 25 & $1.2\times 10^5$ & $3.6\times 10^{15}$ & $1\times 10^{16}$ & $7.6\times 10^{48}$ & $\pi/6$
& 3 & 27 & $150$ & $1.85\times 10^5$ & $3.8\times 10^{15}$ & $4\times 10^{49}$\\
& 4 & 26 & $7.3\times 10^4$ & $7.8\times 10^{15}$ & $1\times 10^{16}$ & $1.04\times 10^{49}$ & $\pi/6$
& 4 & 31 & $150$ & $1.55\times 10^5$ & $2.8\times 10^{15}$ & $8\times 10^{49}$\\
& 5 & 36 & $5\times 10^4$ & $1\times 10^{16}$ & $1\times 10^{16}$ & $1\times 10^{49}$ & $\pi/6$
& 5 & 38 & $150$ & $1.5\times 10^5$ & $7\times 10^{15}$ & $1.76\times 10^{50}$\\
& 6 & 45 & $6.7\times 10^4$ & $1.1\times 10^{16}$ & $1.3\times 10^{16}$ & $5.6\times 10^{48}$ & $\pi/6$
& 6 & 50 & $150$ & $2\times 10^5$ & $5.8\times 10^{15}$ & $7.6\times 10^{49}$\\
& 7 & 56 & $1.15\times 10^5$ & $2.6\times 10^{15}$ & $1\times 10^{16}$ & $2.16\times 10^{49}$ & $\pi/6$
& 7 & 57 & $150$ & $1.4\times 10^5$ & $2.6\times 10^{15}$ & $7.6\times 10^{49}$\\
& 8 & 58 & $6\times 10^4$ & $7.3\times 10^{15}$ & $1\times 10^{16}$ & $1.4\times 10^{49}$ & $\pi/6$
& 8 & 62 & $150$ & $1.2\times 10^5$ & $3.3\times 10^{15}$ & $1.52\times 10^{50}$\\
& 9 & 63 & $6.3\times 10^4$ & $4.9\times 10^{15}$ & $1\times 10^{16}$ & $2.8\times 10^{49}$ & $\pi/6$
& 9 & 66 & $150$ & $1.05\times 10^5$ & $2.6\times 10^{15}$ & $1.72\times 10^{50}$\\
& 10 & 68 & $8.1\times 10^4$ & $2.6\times 10^{15}$ & $1\times 10^{16}$ & $4.6\times 10^{49}$ & $\pi/6$
& 10 & 69 & $150$ & $1.1\times 10^5$ & $2.5\times 10^{15}$ & $1.04\times 10^{50}$\\
& 11 & 71 & $4.8\times 10^4$ & $1\times 10^{16}$ & $1.6\times 10^{16}$ & $8\times 10^{48}$ & $\pi/6$
& 11 & 73 & $150$ & $1.45\times 10^5$ & $9\times 10^{15}$ & $1.12\times 10^{50}$\\
  \hline
171227A & 1 & 8 & $1.75\times 10^5$ & $5.2\times 10^{15}$ & $1\times 10^{16}$ & $2.4\times 10^{49}$ & $\pi/6$
& 1 & 13 & $50$ & $2.3\times 10^5$ & $5\times 10^{14}$ & $5\times 10^{50}$\\
 & 2 & 17 & $3\times 10^5$ & $1\times 10^{15}$ & $3.5\times 10^{15}$ & $1.08\times 10^{50}$ & $\pi/6$
& 2 & 26 & $50$ & $7.8\times 10^4$ & --- & $2.8\times 10^{50}$\\
& 3 & 18.5 & $2.4\times 10^5$ & $1\times 10^{15}$ & $2\times 10^{15}$ & $1.24\times 10^{50}$ & $\pi/6$ &&&&&&\\
& 4 & 19.5 & $2.5\times 10^5$ & $1\times 10^{15}$ & $3.2\times 10^{15}$ & $1.24\times 10^{50}$ & $\pi/6$ &&&&&&\\
& 5 & 22 & $1.9\times 10^5$ & $2.6\times 10^{15}$ & $1\times 10^{16}$ & $8\times 10^{49}$ & $\pi/6$ &&&&&&\\
& 6 & 30 & $9\times 10^4$ & $3.8\times 10^{15}$ & $1\times 10^{16}$ & $7.2\times 10^{49}$ & $\pi/6$ &&&&&&\\
& 7 & 36 & $8\times 10^4$ & $3\times 10^{15}$ & $1\times 10^{16}$ & $9.6\times 10^{49}$ & $\pi/6$ &&&&&&\\
  \hline
180120A & 1 & -2 & $9.4\times 10^4$ & $2.9\times 10^{15}$ & $1\times 10^{16}$ & $1.32\times 10^{50}$ & $\pi/6$
& 1 & -0.5 & $150$ & $1.4\times 10^5$ & $3\times 10^{15}$ & $4.8\times 10^{50}$\\
 & 2 & -1 & $1.4\times 10^5$ & $1.2\times 10^{15}$ & $1\times 10^{16}$ & $2.72\times 10^{50}$ & $\pi/6$
& 2 & 0.5 & $150$ & $9\times 10^4$ & $1.2\times 10^{15}$ & $5.6\times 10^{50}$\\
 & 3 & 4 & $1.3\times 10^5$ & $1.4\times 10^{15}$ & $1\times 10^{16}$ & $2\times 10^{50}$ & $\pi/6$
& 3 & 6 & $150$ & $9.6\times 10^4$ & $1.3\times 10^{15}$ & $3.2\times 10^{50}$\\
 & 4 & 9 & $7.7\times 10^4$ & $5\times 10^{15}$ & $1\times 10^{16}$ & $7.6\times 10^{49}$ & $\pi/6$
& 4 & 12 & $150$ & $1.7\times 10^5$ & $4.4\times 10^{15}$ & $7.6\times 10^{50}$\\
 & 5 & 12 & $1.2\times 10^5$ & $4\times 10^{15}$ & $1\times 10^{16}$ & $3.28\times 10^{49}$ & $\pi/6$
& 5 & 21 & $150$ & $6.3\times 10^4$ & $1\times 10^{15}$ & $4.8\times 10^{50}$\\
& 6 & 18 & $9.5\times 10^4$ & $1.9\times 10^{15}$ & $1\times 10^{16}$ & $2.28\times 10^{50}$ & $\pi/6$ &&&&&&\\
  \hline
180427A & 1 & -1.5 & $1.9\times 10^3$ & $7\times 10^{16}$ & --- & $3.6\times 10^{49}$ & $\pi/6$
& 1 & 0 & $30$ & $3.3\times 10^4$ & --- & $4.4\times 10^{51}$\\
  \hline
180806A & 1 & -1 & $1.2\times 10^4$ & $1.6\times 10^{16}$ & --- & $3.6\times 10^{48}$ & $\pi/6$
& 1 & 0 & $60$ & $7.4\times 10^4$ & --- & $7.6\times 10^{49}$\\
 & 2 & 1 & $1.9\times 10^5$ & $1\times 10^{15}$ & $8\times 10^{15}$ & $2.2\times 10^{49}$ & $\pi/6$
& 2 & 2 & $60$ & $9.3\times 10^4$ & $2\times 10^{14}$ & $1.08\times 10^{50}$\\
 & 3 & 2 & $1.9\times 10^5$ & $1.3\times 10^{15}$ & $1\times 10^{16}$ & $2.8\times 10^{49}$ & $\pi/6$
& 3 & 4 & $60$ & $9\times 10^4$ & $2.3\times 10^{14}$ & $1.08\times 10^{50}$\\
& 4 & 5 & $1.9\times 10^5$ & $1\times 10^{15}$ & $3.6\times 10^{15}$ & $1.92\times 10^{49}$ & $\pi/6$
& 4 & 6 & $60$ & $1.1\times 10^5$ & $1.5\times 10^{14}$ & $4.4\times 10^{49}$\\
& 5 & 7 & $2.5\times 10^5$ & $1\times 10^{15}$ & $4.7\times 10^{15}$ & $1.04\times 10^{49}$ & $\pi/6$ &&&&&&\\
  \hline
180914A & 1 & 8 & $1.3\times 10^5$ & $1\times 10^{15}$ & $5\times 10^{15}$ & $5.2\times 10^{49}$ & $\pi/6$
& 1 & 9 & $150$ & $9.2\times 10^4$ & $1\times 10^{15}$ & $1\times 10^{50}$\\
 & 2 & 9.5 & $1.35\times 10^5$ & $1.3\times 10^{15}$ & $9.5\times 10^{15}$ & $7.2\times 10^{49}$ & $\pi/6$
& 2 & 11 & $150$ & $9.4\times 10^4$ & $1.2\times 10^{15}$ & $1.48\times 10^{50}$\\
 & 3 & 14 & $1.5\times 10^5$ & $1\times 10^{15}$ & $5.9\times 10^{15}$ & $6.4\times 10^{49}$ & $\pi/6$
& 3 & 14 & $150$ & $1.2\times 10^5$ & $1.6\times 10^{15}$ & $9.2\times 10^{49}$\\
& 4 & 16.5 & $1.1\times 10^5$ & $1\times 10^{15}$ & $4.8\times 10^{15}$ & $1.16\times 10^{50}$ & $\pi/6$
& 4 & 17.3 & $150$ & $7.9\times 10^4$ & $9.1\times 10^{14}$ & $1.36\times 10^{50}$\\
& 5 & 20 & $1.1\times 10^5$ & $1\times 10^{15}$ & $4\times 10^{15}$ & $1\times 10^{50}$ & $\pi/6$
& 5 & 20.5 & $150$ & $8.7\times 10^4$ & $1\times 10^{15}$ & $1.28\times 10^{50}$\\
& 6 & 21 & $1.2\times 10^5$ & $1\times 10^{15}$ & $5.2\times 10^{15}$ & $8\times 10^{49}$ & $\pi/6$
& 6 & 21.5 & $150$ & $1\times 10^5$ & $1.3\times 10^{15}$ & $9.2\times 10^{49}$\\
& 7 & 24 & $9.8\times 10^4$ & $2\times 10^{15}$ & $1\times 10^{16}$ & $4\times 10^{49}$ & $\pi/6$
& 7 & 25 & $150$ & $1.1\times 10^5$ & $2.5\times 10^{15}$ & $1.16\times 10^{50}$\\
& 8 & 28.5 & $9.5\times 10^4$ & $2.6\times 10^{15}$ & $1\times 10^{16}$ & $3.72\times 10^{49}$ & $\pi/6$
& 8 & 31 & $150$ & $9.7\times 10^4$ & $1.8\times 10^{15}$ & $1.48\times 10^{50}$\\
& 9 & 60 & $6.5\times 10^4$ & $8.7\times 10^{15}$ & $1\times 10^{16}$ & $6.8\times 10^{48}$ & $\pi/6$
& 9 & 66 & $150$ & $1.5\times 10^5$ & $4.2\times 10^{15}$ & $9.6\times 10^{49}$\\
& 10 & 72.5 & $9.3\times 10^4$ & $3.7\times 10^{15}$ & $1\times 10^{16}$ & $9.2\times 10^{48}$ & $\pi/6$
& 10 & 78 & $150$ & $9.7\times 10^4$ & $1.1\times 10^{15}$ & $5.2\times 10^{49}$\\
& 11 & 81 & $6.2\times 10^4$ & $8.2\times 10^{15}$ & $1\times 10^{16}$ & $1.28\times 10^{49}$ & $\pi/6$
& 11 & 88 & $150$ & $1.3\times 10^5$ & $3.3\times 10^{15}$ & $1.72\times 10^{50}$\\
& 12 & 123 & $7.7\times 10^4$ & $5\times 10^{15}$ & $1\times 10^{16}$ & $1.24\times 10^{49}$ & $\pi/6$
& 12 & 127 & $150$ & $1.4\times 10^5$ & $3.1\times 10^{15}$ & $9.2\times 10^{49}$\\
  \hline
190530A & 1 & 3.5 & $2.4\times 10^5$ & $2.7\times 10^{15}$ & $1\times 10^{16}$ & $1.56\times 10^{50}$ & $\pi/6$
& 1 & 7 & $150$ & $2.3\times 10^5$ & $1.4\times 10^{15}$ & $7.6\times 10^{50}$\\
 & 2 & 9 & $2.3\times 10^5$ & $1.9\times 10^{15}$ & $1\times 10^{16}$ & $1.24\times 10^{50}$ & $\pi/6$
& 2 & 11.5 & $150$ & $1.9\times 10^5$ & $1.4\times 10^{15}$ & $4.4\times 10^{50}$\\
 & 3 & 13 & $3.1\times 10^5$ & $1\times 10^{15}$ & $6.2\times 10^{15}$ & $3.6\times 10^{50}$ & $\pi/6$
& 3 & 14 & $150$ & $2\times 10^5$ & $9\times 10^{14}$ & $6\times 10^{50}$\\
& 4 & 15 & $2.5\times 10^5$ & $1\times 10^{15}$ & $2.8\times 10^{15}$ & $2.72\times 10^{50}$ & $\pi/6$
& 4 & 16 & $150$ & $1.9\times 10^5$ & $4.2\times 10^{14}$ & $2.68\times 10^{50}$\\
& 5 & 16 & $2.5\times 10^5$ & $1\times 10^{15}$ & $2\times 10^{15}$ & $1.48\times 10^{50}$ & $\pi/6$
& 5 & 16.5 & $150$ & $2.2\times 10^5$ & $6.3\times 10^{14}$ & $1.96\times 10^{50}$\\
& 6 & 16.5 & $2.4\times 10^5$ & $1\times 10^{15}$ & $4.2\times 10^{15}$ & $4\times 10^{50}$ & $\pi/6$
& 6 & 17.5 & $150$ & $1.6\times 10^5$ & $6.2\times 10^{14}$ & $4.4\times 10^{50}$\\
  \hline
200311A & 1 & 0 & $1.9\times 10^5$ & $1\times 10^{15}$ & $1.8\times 10^{15}$ & $5.6\times 10^{49}$ & $\pi/6$
& 1 & 0.5 & $150$ & $1.45\times 10^5$ & $5.7\times 10^{14}$ & $1.04\times 10^{50}$\\
 & 2 & 1 & $2.8\times 10^5$ & $1\times 10^{15}$ & $2.5\times 10^{15}$ & $2.48\times 10^{49}$ & $\pi/6$
& 2 & 1.5 & $150$ & $2.1\times 10^5$ & $8.3\times 10^{14}$ & $4.8\times 10^{49}$\\
 & 3 & 2 & $4.7\times 10^5$ & $1\times 10^{15}$ & $6.3\times 10^{15}$ & $8.4\times 10^{48}$ & $\pi/6$
& 3 & 3 & $150$ & $3.5\times 10^5$ & $1.2\times 10^{15}$ & $9.2\times 10^{48}$\\
& 4 & 3 & $5\times 10^5$ & $1.1\times 10^{15}$ & $1\times 10^{16}$ & $8.8\times 10^{48}$ & $\pi/6$
& 4 & 5 & $150$ & $3.1\times 10^5$ & $1.4\times 10^{15}$ & $1.72\times 10^{49}$\\
& 5 & 5.5 & $4\times 10^5$ & $1\times 10^{15}$ & $5\times 10^{15}$ & $9.6\times 10^{48}$ & $\pi/6$
& 5 & 6 & $150$ & $4.2\times 10^5$ & $1.7\times 10^{15}$ & $9.6\times 10^{48}$\\
& 6 & 7 & $4.4\times 10^5$ & $1\times 10^{15}$ & $2.3\times 10^{15}$ & $4\times 10^{48}$ & $\pi/6$
& 6 & 9.5 & $150$ & $5.5\times 10^5$ & $3\times 10^{15}$ & $1.08\times 10^{49}$\\
& 7 & 7.5 & $4\times 10^5$ & $1.2\times 10^{15}$ & $1\times 10^{16}$ & $9.6\times 10^{48}$ & $\pi/6$
& 7 & 20 & $150$ & $2\times 10^5$ & $4.3\times 10^{15}$ & $8.4\times 10^{49}$\\
& 8 & 8 & $3.8\times 10^5$ & $2.8\times 10^{15}$ & $1\times 10^{16}$ & $3\times 10^{48}$ & $\pi/6$
& 8 & 30 & $150$ & $6.5\times 10^5$ & $7.5\times 10^{15}$ & $7.6\times 10^{48}$\\
& 9 & 20 & $1.1\times 10^5$ & $4\times 10^{15}$ & $1\times 10^{16}$ & $1.88\times 10^{49}$ & $\pi/6$ &&&&&&\\
& 10 & 30 & $2.15\times 10^5$ & $7.5\times 10^{15}$ & $1\times 10^{16}$ & $1\times 10^{48}$ & $\pi/6$ &&&&&&\\
  \hline
200412A & 1 & -0.5 & $4.8\times 10^4$ & $1\times 10^{15}$ & $6\times 10^{14}$ & $8.4\times 10^{50}$ & $\pi/6$
& 1 & -0.2 & $150$ & $4.2\times 10^4$ & $3.5\times 10^{14}$ & $1.4\times 10^{51}$\\
 & 2 & 0.4 & $4.5\times 10^4$ & $1\times 10^{15}$ & $2.2\times 10^{14}$ & $6.4\times 10^{50}$ & $\pi/6$
& 2 & 0.5 & $150$ & $4.6\times 10^4$ & $2\times 10^{14}$ & $6\times 10^{50}$\\
 & 3 & 0.7 & $5.1\times 10^4$ & $1\times 10^{15}$ & $5.5\times 10^{14}$ & $6\times 10^{50}$ & $\pi/6$
& 3 & 0.9 & $150$ & $4.7\times 10^4$ & $3.6\times 10^{14}$ & $8\times 10^{50}$\\
& 4 & 2 & $8.4\times 10^4$ & $1\times 10^{15}$ & $2.3\times 10^{15}$ & $6\times 10^{50}$ & $\pi/6$
& 4 & 2.8 & $150$ & $5.8\times 10^4$ & $5\times 10^{14}$ & $1\times 10^{51}$\\
& 5 & 3.5 & $6.4\times 10^4$ & $1\times 10^{15}$ & $1.9\times 10^{14}$ & $4.4\times 10^{50}$ & $\pi/6$
& 5 & 3.5 & $150$ & $6.6\times 10^4$ & $2.4\times 10^{14}$ & $4.8\times 10^{50}$\\
& 6 & 4 & $4.7\times 10^4$ & $1\times 10^{15}$ & $3.7\times 10^{14}$ & $7.2\times 10^{50}$ & $\pi/6$
& 6 & 4.3 & $150$ & $3.5\times 10^4$ & $2\times 10^{14}$ & $1.64\times 10^{51}$\\
& 7 & 4.5 & $5.6\times 10^4$ & $1\times 10^{15}$ & $4.2\times 10^{14}$ & $2\times 10^{50}$ & $\pi/6$
& 7 & 4.7 & $150$ & $6.3\times 10^4$ & $4.5\times 10^{14}$ & $2.32\times 10^{50}$\\
& 8 & 8 & $6.9\times 10^4$ & $1\times 10^{15}$ & $1.3\times 10^{15}$ & $4.4\times 10^{50}$ & $\pi/6$
& 8 & 8.6 & $150$ & $5.1\times 10^4$ & $3.8\times 10^{14}$ & $7.6\times 10^{50}$\\
& 9 & 11 & $7.4\times 10^4$ & $1\times 10^{15}$ & $8.6\times 10^{14}$ & $4\times 10^{50}$ & $\pi/6$
& 9 & 11.7 & $150$ & $4.8\times 10^4$ & $2\times 10^{14}$ & $1.04\times 10^{51}$\\
& 10 & 12 & $5\times 10^4$ & $1\times 10^{15}$ & $3.6\times 10^{14}$ & $2.8\times 10^{50}$ & $\pi/6$
& 10 & 12 & $150$ & $5.1\times 10^4$ & $3.5\times 10^{14}$ & $3.6\times 10^{50}$\\
\end{longtable}
\tablefoot{The parameters in the left (right) column correspond to the aligned-fields (toroidal-fields) case. For the aligned-fields case, we take $\Gamma_{0} = 250$. For the toroidal-fields case, we take $r_0 = 1\times 10^{15}$ cm.\\
\tablefootmark{a}{The variation of electron Lorentz factor is a single power law with $g=-1$ for the aligned-fields case, while it is $g=-1.5$ for the toroidal-fields case.}
}
\end{landscape}

\end{appendix}
\end{document}